\documentclass[11pt]{article}
\usepackage[bottom=1in,left=1in, right=1in, top=1in]{geometry}
\usepackage[utf8]{inputenc}
\usepackage[T1]{fontenc}
\usepackage{authblk}
\usepackage{amsmath}
\usepackage{amssymb}
\usepackage{amsfonts}
\usepackage{amsthm}
\usepackage{mathtools}
\usepackage{bbm}
\usepackage{nicefrac}
\usepackage{algorithm}
\usepackage[noend]{algpseudocode}
\usepackage{tabularx}
\usepackage{enumitem}
\usepackage{graphicx}
\usepackage{subfig}          
\usepackage{float}           
\usepackage{placeins}     

\makeatletter
\g@addto@macro\@floatboxreset\centering
\makeatother
\usepackage{booktabs}
\usepackage{xcolor}
\usepackage{tikz}
\usetikzlibrary{calc}
\usepackage{microtype}
\usepackage{url}
\usepackage{xurl}          
\usepackage{cite}
\usepackage{hyperref}

\hypersetup{
	colorlinks=true, 
	linktoc=all,     
	linkcolor=blue,  
	citecolor = blue
}

\usepackage{titlesec}
\titlespacing{\paragraph}{0pt}{5pt}{5pt}

\providecommand{\Description}[2][]{}

\newtheorem{theorem}{Theorem}
\newtheorem{lemma}[theorem]{Lemma}
\newtheorem{observation}{Observation}
\newtheorem{property}{Property}

\newcommand{\edgegap}{0.35}

\newcommand{\problem} [1] {\textnormal{\textsc{#1}}}

\def\minecc{\problem{MinECC}}

\def\SPECC{\problem{SPECC}}

\def\RECC{\problem{RobustECC}}

\definecolor{red}{HTML}{CF5555}
\definecolor{blue}{HTML}{4C87BB}

\definecolor{purple}{HTML}{8050A3}
\definecolor{green}{HTML}{329653}
\definecolor{yellow}{HTML}{D9AD00}

\newcommand{\drawvertices}{
\node[vertex] at (110:1.5) {};
\node[vertex] at (130:1.5) {};
\node[vertex] at (170:1.5) {};
\node[vertex] at (190:1.5) {};
\node[vertex] at (-10:1.5) {};
\node[vertex] at (10:1.5) {};
\node[vertex] at (50:1.5) {};
\node[vertex] at (70:1.5) {};
\node[vertex] at (230:1.5) {};
\node[vertex] at (250:1.5) {};
\node[vertex] at (290:1.5) {};
\node[vertex] at (310:1.5) {};
}

\title{Edge-Colored Clustering with Soft-Penalized Node Deletions}

\author[1]{Abhinav Sairam}
\author[2]{Srikar Yalam}
\author[2]{Nate Veldt}

\affil[1]{Department of Mathematics, Texas A\&M University}
\affil[2]{Department of Computer Science and Engineering, Texas A\&M University}
\date{}

\begin{document}

\maketitle
\begin{abstract}
Given an edge-colored hypergraph, the edge-colored clustering (ECC) problem seeks to delete a minimum number of edges in order to remove conflicts, i.e., overlapping edges of different colors.
This provides a framework for clustering data based on categorical group interactions.
We present algorithms for a new generalization of the problem where one is additionally allowed to delete nodes to resolve conflicts. The cost of node deletion is governed by a tunable parameter, and we provide constant-factor approximation algorithms that work for any choice of parameter. We also show how to obtain approximate solutions for a sequence of parameters while ensuring solutions across different parameters are correlated by a natural nesting property. In particular, if a node is deleted for one value of the penalty parameter, it will also be deleted for all smaller penalty parameter. We demonstrate the performance and utility of our algorithms on hypergraph datasets where node deletions are especially useful for identifying outliers, including results on a new edge-colored hypergraph with over 1 million hyperedges, encoding posts on Stackexchange.
\end{abstract}

\section{Introduction}
Edge-Colored Clustering (ECC) is a combinatorial framework for clustering and classification in datasets where data points participate in overlapping multiway interactions of different categories~\cite{amburg2020clustering}. For this problem, data is modeled as a hypergraph where nodes represent data points, hyperedges represent group interactions between nodes, and colors indicate a category of interaction. Among other examples, this can be used to capture co-authored articles on the Microsoft Academic Graph (where colors represents article topics)~\cite{amburg2020clustering}, groups of vacation rentals on Trivago visited during the same browsing session (where color represents location of the browsing session)~\cite{veldt2023optimal}, or groups of reviewers who have reviewed the same product on Amazon or restaurant on Yelp (where color represents product/restaurant category)~\cite{amburg2022Diverse}. The goal of ECC is then to color nodes in a way that avoids unsatisfied edges---edges where even a single node is given a color different from the edge's color. In the past several years there has been a significant amount of interest in the data mining and machine learning community on developing and applying approximation algorithms and parameterized complexity results for variants of ECC~\cite{amburg2020clustering,amburg2022Diverse,veldt2023optimal,lee2025improvedalgorithmsoverlappingrobust,crane2025edge,han2026improved,srinivasan2026edge,kellerhals2023parameterized,CraneEtAl2024Overlapping}.

In standard ECC, every node must be given a single color. However, in many applications is it natural for certain data objects to be outliers that do not belong to any class. For example, an author may be interdisciplinary and not belong to multiple fields, or an online reviewer may tend to review restaurants or products from a wide variety of different cuisine or product categories. 
This motivated Crane et al.~\cite{CraneEtAl2024Overlapping} to develop recent generalizations of ECC where some nodes can belong to multiple categories (overlapping ECC), or may be treated as complete outliers and deleted (\RECC). For each of these variants, the problem input is defined by an edge-colored hypergraph along with an integer budget that strictly bounds the amount of overlap or node deletions. For \RECC{} specifically, the goal is to minimize the number of unsatisfied edges after deleting a fixed number of nodes. Prior work has shown how to develop approximation algorithms whose approximation factor is proportional to the budget, as well as bicriteria approximation algorithms (which approximately minimize the objective while approximately satisfying the constraint)~\cite{crane2025edge,lee2025improvedalgorithmsoverlappingrobust}.

\paragraph{Our work: ECC with soft-penalized node deletions.}
\begin{figure}[t]
  \centering
  \subfloat[Original hypergraph]
    {
    \resizebox{0.25\linewidth}{!}{%
    \begin{tikzpicture}
    [
        line width = 1.5pt,
        vertex/.style={circle, draw, fill=white}
    ]

    \draw[purple, rotate=120] (0.8,0) ellipse (1.15 and 0.68);
    \draw[purple, rotate=180] (0.8,0) ellipse (1.15 and 0.68);
    \draw[green, rotate=0] (0.8,0) ellipse (1.15 and 0.68);
    \draw[green, rotate=60] (0.8,0) ellipse (1.15 and 0.68);
    \draw[yellow, rotate=240] (0.8,0) ellipse (1.15 and 0.68);
    \draw[yellow, rotate=300] (0.8,0) ellipse (1.15 and 0.68);

    \node[vertex] at (0:0) {$v$};
    \drawvertices
    \end{tikzpicture}
    }
    }
    \hfill
    \subfloat[\minecc{} solution]
    {
    \resizebox{0.25\linewidth}{!}{%
    \begin{tikzpicture}
    [
        line width = 1.5pt,
        vertex/.style={circle, draw, fill=white}
    ]

    \draw[purple, rotate=120] (0.8,0) ellipse (1.15 and 0.68);
    \draw[purple, rotate=180] (0.8,0) ellipse (1.15 and 0.68);
    \draw[green, dashed, rotate=0] (0.8,0) ellipse (1.15 and 0.68);
    \draw[green, dashed, rotate=60] (0.8,0) ellipse (1.15 and 0.68);
    \draw[yellow, dashed, rotate=240] (0.8,0) ellipse (1.15 and 0.68);
    \draw[yellow, dashed, rotate=300] (0.8,0) ellipse (1.15 and 0.68);

    \node[vertex,fill=purple] at (0:0) {$v$};
    \node[vertex,fill=purple] at (110:1.5) {};
    \node[vertex,fill=purple] at (130:1.5) {};
    \node[vertex,fill=purple] at (170:1.5) {};
    \node[vertex,fill=purple] at (190:1.5) {};
    \node[vertex,fill=green] at (-10:1.5) {};
    \node[vertex,fill=green] at (10:1.5) {};
    \node[vertex,fill=green] at (50:1.5) {};
    \node[vertex,fill=green] at (70:1.5) {};
    \node[vertex,fill=yellow] at (230:1.5) {};
    \node[vertex,fill=yellow] at (250:1.5) {};
    \node[vertex,fill=yellow] at (290:1.5) {};
    \node[vertex,fill=yellow] at (310:1.5) {};
    \end{tikzpicture}
    }}
    \hfill
    \subfloat[\SPECC{} solution]
    {
    \resizebox{0.25\linewidth}{!}{%
    \begin{tikzpicture}
    [
        line width = 1.5pt,
        vertex/.style={circle, draw, fill=white}
    ]

    \draw[purple, rotate=120] (1.48,0) ellipse (0.40 and 0.62);
    \draw[purple, rotate=180] (1.48,0) ellipse (0.40 and 0.62);
    \draw[green, rotate=0] (1.48,0) ellipse (0.40 and 0.62);
    \draw[green, rotate=60] (1.48,0) ellipse (0.40 and 0.62);
    \draw[yellow, rotate=240] (1.48,0) ellipse (0.40 and 0.62);
    \draw[yellow, rotate=300] (1.48,0) ellipse (0.40 and 0.62);

    \node[vertex,fill=purple] at (110:1.5) {};
    \node[vertex,fill=purple] at (130:1.5) {};
    \node[vertex,fill=purple] at (170:1.5) {};
    \node[vertex,fill=purple] at (190:1.5) {};
    \node[vertex,fill=green] at (-10:1.5) {};
    \node[vertex,fill=green] at (10:1.5) {};
    \node[vertex,fill=green] at (50:1.5) {};
    \node[vertex,fill=green] at (70:1.5) {};
    \node[vertex,fill=yellow] at (230:1.5) {};
    \node[vertex,fill=yellow] at (250:1.5) {};
    \node[vertex,fill=yellow] at (290:1.5) {};
    \node[vertex,fill=yellow] at (310:1.5) {};

    \end{tikzpicture}
    }}
  \caption{(a) A small 3-colored hypergraph, with unit weights on nodes and edges. (b) The standard \minecc{} solution deletes four edges. (c) For $\gamma < 4$, the \SPECC{} solution simply deletes the center node, correctly identifying it as an outlier that should not be viewed as having a single correct color assignment.}
  \Description{insert}
  \label{fig:specc}
\end{figure}
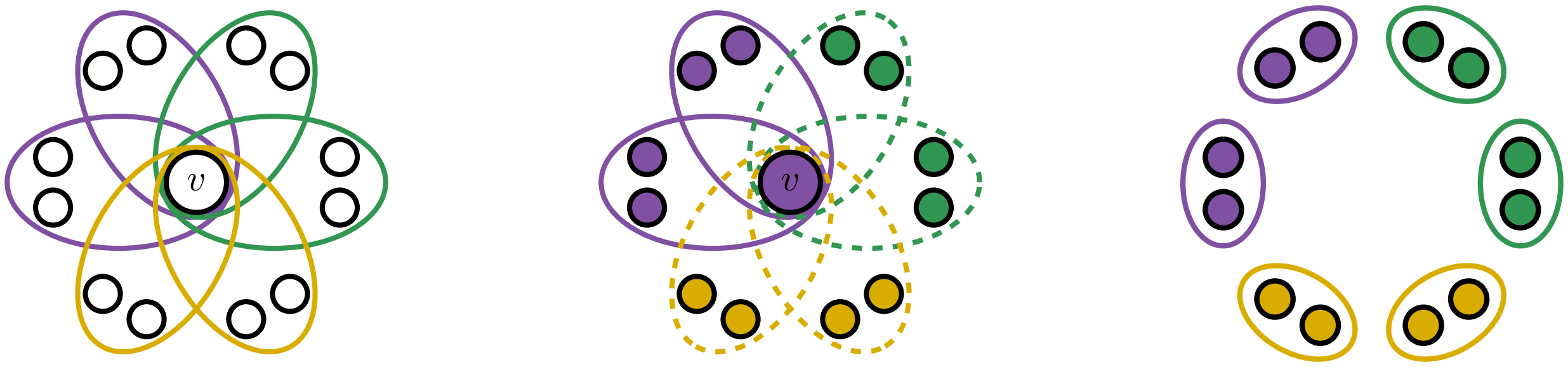
Our work builds on previous research on ECC with node outliers~\cite{CraneEtAl2024Overlapping,lee2025improvedalgorithmsoverlappingrobust} in several directions. We first of all present a new variant of ECC in which node deletions are not controlled by a fixed budget (as in \RECC) but by a tunable soft penalty parameter $\gamma \geq 0$. We call our objective Soft-Penalized ECC (\SPECC). When $\gamma = \infty$, the cost for deleting a node is infinite and the problem reduces to standard ECC, but for smaller parameter values it becomes possible to delete nodes to resolve conflicts (i.e., pairs of different-colored edges that overlap). See Figure~\ref{fig:specc} for illustration.
Our objective additionally differs from \RECC{} in that we associated an arbitrary individual penalty weight to each node (independent of the global parameter $\gamma$), rather than simply bounding or penalizing the exact \emph{number} of deleted nodes. In practice, this allows us to better identify different \emph{types} of node outliers that exist in the same dataset. After introducing the \SPECC{} problem, we show how to develop constant-factor approximation algorithms for any choice of node weights and parameter $\gamma$. In particular, we prove that (unlike \RECC{}), we can develop approximation algorithms via reduction to hypergraph vertex cover and node-weighted multiway cut. Explicitly performing these reductions can be expensive, so we then develop a more scalable $(2-2/k)$-approximation based on rounding a linear programming (LP) relaxation for our problem. We prove this LP relaxation provides a tighter lower bound than the one obtained by reducing to vertex cover or node-weighted multiway cut.

Another key algorithmic contribution of our work is to develop algorithms that find approximate solutions for an entire \emph{range} of $\gamma$ values, while ensuring that the solutions are meaningfully correlated across different parameters via a useful nesting property. 
Specifically, we design an algorithm which, when given a range of parameters $\{\gamma_1, \gamma_2, \hdots, \gamma_k\}$, finds a $3$-approximate solution for each $\gamma_i$ while ensuring that the set of edges that is deleted is strictly growing as $\gamma$ increases, and the set of nodes that are deleted is strictly shrinking as $\gamma$ increases. Although this property is intuitive and natural, we show that LP-rounding algorithms that round solutions for different $\gamma$ values separately do not satisfy it. 
In addition, our new combinatorial algorithm is very fast; for each $\gamma_i$ it takes linear time in terms of the size of the hypergraph.

In additional to our theoretical results, we provide in-depth empirical results on several real-world datasets where node outliers are particularly relevant. We first of all demonstrate the quality and speed of our algorithms across a range of different hypergraph datasets. This includes results for a new edge-colored hypergraph dataset which we introduce, where nodes are registered users on Stackexchange.com, hyperedges are users that interact with a specific question (by asking, answering, or commenting), and colors indicate the sub-site associated with the question (e.g., \emph{math.stackexchange.com} or \emph{bioacoustics.stackexchange.com}). We also provided a detailed case study on the Cooking hypergraph, where nodes are ingredients, hyperedges are recipes, and colors indicate cuisine. In particular, we show how varying node weights as well as the global parameter $\gamma$ allows us to identify different types of outlier ingredients. Our case studies demonstrate the benefits of having algorithms which (1) provide correlated (nested) solutions as $\gamma$ varies, and (2) allow node-specific penalties for deletion.

\section{Preliminaries and related work}
An instance of Edge-colored Clustering (ECC) is given by a hypergraph $H = (V, E, C, \ell)$; $C = [k] = \{1,2, \hdots, k\}$ is a set of ($k$) colors and $\ell \colon E \rightarrow C$ is a map from (hyper)edges to colors. The rank of the hypergraph is $r = \max_e |e|$. Each edge $e \in E$ can additionally be associated with a non-negative weight $w(e) \geq 0$ 
(the unweighted case is when $w(e) = 1$ for every $e \in E$). The degree of $v \in V$ (the number of incident edges) is denoted by $d_v$. The \emph{color-degree} of $v$ is the number of colors among its incident edges.

The standard minimum Edge-Colored Clustering (\minecc{}) objective seeks to build a map from nodes to colors $\lambda \colon V \rightarrow C$ that minimizes the weight of unsatisfied edges. An edge $e \in E$ is unsatisfied if there exists any $v \in e$ such that $\lambda(v) \neq \ell(e)$. \minecc{} can equivalently be formulated as an edge-deletion problem where the goal is to remove a minimum weight set of edges to resolve \emph{conflicts}, which we also call \emph{bad edge pairs}. A conflict is defined to be a pair of edges $e, f \in E$ such that $e \cap f \neq \varnothing$ and $\ell(e) \neq \ell(f)$. We will also use the term \emph{bad triplet} to denote a triplet $\{v, e, f\}$ where $v \in e\cap f$ and $\ell(e) \neq \ell(f)$. 
\minecc{} is also equivalent to deleting a minimum weight set of edges so that nodes in the remaining hypergraph have color-degree equal to 1. 
Formally, if we use $D_E \subseteq E$ to denote a set of edges to delete, the standard \minecc{} objective is to solve
\begin{equation}
    \label{eq:minecc}
    {\min \quad \sum_{e \in D_E} w(e) \quad \text{ s.t.\ $E\setminus D_E$ contains no conflicts.}}
\end{equation}

\paragraph{Prior algorithmic results and variants.} Amburg et al.~\cite{amburg2020clustering} proved \minecc{} in hypergraphs is NP-hard and gave a $\min\{2 - 1/k, 2- 1/(r+1)\}$ approximation based on rounding an LP relaxation. Veldt~\cite{veldt2023optimal} gave a $\min\{2 - 2/k, 2- 2/(r+1)\}$-approximation via LP rounding, with a matching integrality gap. Veldt~\cite{veldt2023optimal} also designed a faster combinatorial 2-approximation based on reduction to vertex cover. Recently, Han and Veldt~\cite{han2026improved} gave a combinatorial and nearly-linear time $(2-2/k)$-approximation. Many papers have also explored the approximability of assigning colors to nodes to maximize satisfied edges. This includes extensive work for the case of graphs~\cite{angel2016clustering,ageev2014improved,ageev20200,alhamdan2019approximability,crane2025edge}, and some work for hypergraphs~\cite{crane2025edge,srinivasan2026edge}.


\paragraph{Robust ECC and other generalizations.} 
Our work is most closely related to that of Crane et al.~\cite{CraneEtAl2024Overlapping} on \RECC{}.
For \RECC{}, one is given a budget $b$ on the number of nodes to delete before minimizing unsatisfied edges. They gave a greedy $r$ approximation as well as bicriteria approximations based on LP-rounding. Recently, Lee et al.~\cite{lee2025improvedalgorithmsoverlappingrobust} provided faster algorithms for \RECC{} via the primal-dual method, including a $2(b+1)$-approximation and improved bicriteria approximation factors. Other generalizations of ECC include variants with overlapping clusters~\cite{CraneEtAl2024Overlapping,lee2025improvedalgorithmsoverlappingrobust}, color-balanced variants~\cite{crane2025edge}, and variants that incorporate notions of diversity and experience in color-assignment~\cite{amburg2022Diverse}.

\section{Soft-penalized Edge-Colored Clustering}
\label{sec:specc}
We now introduced a new generalization of ECC which provides an alternative approach for handling node outliers. As in standard ECC, the input to the problem is an edge-colored hypergraph $H = (V,E,C, \ell)$, where $w(e) \geq 0$ is a weight associated with $e \in E$. For our new problem, we extend the weight function to nodes, so that $w(v) \geq 0$ is a weight for node $v \in V$. Additionally, we define a global parameter $\gamma \geq 0$ which can be varied to implicitly control the penalty for deleting nodes. Given this input, the Soft-Penalzed Edge Colored Clustering problem (\SPECC) seeks a minimum-cost set of edges $D_E \subseteq E$ and nodes $D_V \subseteq V$ to delete so that the resulting hypergraph $H' = (V\setminus D_V, E\setminus D_E)$ has no conflicts (overlapping edges of different colors). The cost of a deletion set is:
\begin{equation}
    \label{eq:specc}
    {\textstyle \text{cost}(H, D_E, D_V,\gamma) = \sum_{e \in D_E} w(e) + \gamma\sum_{v \in D_V}w(v).}
\end{equation}
Deleting a node means removing it from all of its incident edges and shrinking the edges accordingly (if an edge ends up with no nodes, it is trivially satisfied). As a generalization of ECC, this is NP-hard. Figure~\ref{fig:specc} provides an illustration on a small hypergraph, where nodes and edges have unit weights.

\paragraph{Differences from \RECC{} and resulting consequences.}
The primary difference between \SPECC{} and \RECC{}~\cite{CraneEtAl2024Overlapping} is that \SPECC{} does not include a hard constraint on deleting nodes but rather a tunable soft penalty. As a result, the number and choice of nodes to delete is determined implicitly by optimizing the objective. We will prove that this difference makes \SPECC{} amenable to constant-factor approximations that are independent of $\gamma$ and node weights. By comparison, existing approximations for \RECC{} either depend on the budget for node deletion or violate hard constraints. Another difference is that \SPECC{}  associates an arbitrary weight for node deletion, rather than treating all node deletions as equal. This is minor from the perspective of theoretical results, but we shall see that it can make a significant difference in practical experiments and exploratory data analysis. In particular, we shall see that different node-weighting schemes allow us to identify different types of outlier nodes in practice. Finally, incorporating a tunable parameter $\gamma$ will allow us to consider notions of a nested sequence of \SPECC{} solutions, where the node and edge deletion sets are monotonic in $\gamma$. This is both interesting from a theoretical perspective (see Section~\ref{sec:nested}) and beneficial in empirical analyses.

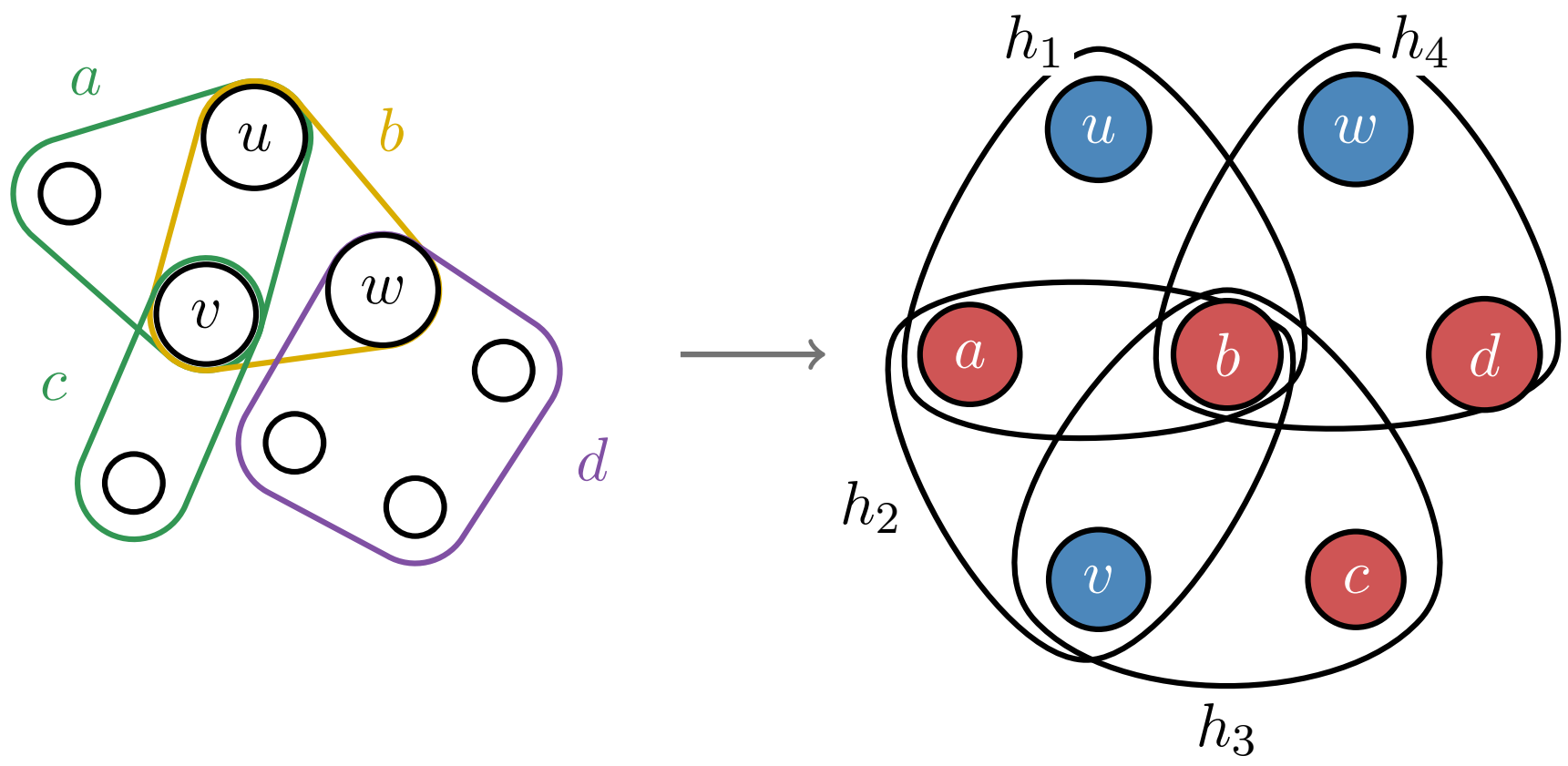
\begin{figure}[t]
  \centering
  {
  \resizebox{.7\linewidth}{!}{%
  \begin{tikzpicture}[
    line width=1pt,
    vertex/.style={circle, draw=black, fill=white}
    ]
\begin{scope}

    \coordinate (p) at (-1.4,1);
    \coordinate (u) at (-0.25,1.35);
    \coordinate (v) at (-0.55,0.25);
    \coordinate (w) at (0.55,0.4);
    \coordinate (q) at (0,-0.55);
    \coordinate (t) at (-1,-0.8);
    \coordinate (r) at (0.75,-0.95);
    \coordinate (s) at (1.3,-0.1);

    \draw[green] ($(p)+(105:\edgegap)$)
      arc[start angle=105, end angle=230, radius=\edgegap]
      -- ($(v)+(230:\edgegap)$)
      arc[start angle=230, end angle=345, radius=\edgegap]
      -- ($(u)+(345:\edgegap)$)
      arc[start angle=345, end angle=465, radius=\edgegap]
      -- cycle;
    \draw[yellow] ($(v)+(165:\edgegap)$)
      arc[start angle=165, end angle=280, radius=\edgegap]
      -- ($(w)+(280:\edgegap)$)
      arc[start angle=280, end angle=400, radius=\edgegap]
      -- ($(u)+(40:\edgegap)$)
      arc[start angle=40, end angle=165, radius=\edgegap]
      -- cycle;
    \draw[green] ($(t)+(155:\edgegap)$)
      arc[start angle=155, end angle=335, radius=\edgegap]
      -- ($(v)+(335:\edgegap)$)
      arc[start angle=335, end angle=515, radius=\edgegap]
      -- cycle;
    \draw[purple] ($(q)+(150:\edgegap)$)
      arc[start angle=150, end angle=240, radius=\edgegap]
      -- ($(r)+(240:\edgegap)$)
      arc[start angle=240, end angle=325, radius=\edgegap]
      -- ($(s)+(325:\edgegap)$)
      arc[start angle=325, end angle=415, radius=\edgegap]
      -- ($(w)+(55:\edgegap)$)
      arc[start angle=55, end angle=150, radius=\edgegap]
      -- cycle;

    \node[vertex] at (p) {};
    \node[vertex] at (u) {$u$};
    \node[vertex] at (v) {$v$};
    \node[vertex] at (w) {$w$};
    \node[vertex] at (q) {};
    \node[vertex] at (t) {};
    \node[vertex] at (r) {};
    \node[vertex] at (s) {};

    \node[green, fill=white, inner sep=1.5pt] at (-1.3,1.7) {$a$};
    \node[yellow, fill=white, inner sep=1.5pt] at (0.6,1.4) {$b$};
    \node[green, fill=white, inner sep=1.5pt] at (-1.5,-0.2) {$c$};
    \node[purple, fill=white, inner sep=1.5pt] at (1.85,-0.65) {$d$};
\end{scope}

\draw[->, black!55] (2.4,0) -- (3.3,0);

\begin{scope}[xshift=5.8cm]

    \coordinate (A) at (-1.6,0);
    \coordinate (B) at (0,0);
    \coordinate (C) at (0.8,-1.4);
    \coordinate (D) at (1.6,0);

    \coordinate (U) at (-0.8,1.4);
    \coordinate (V) at (-0.8,-1.4);
    \coordinate (W) at (0.8,1.4);


    \draw[black] (-1.95,-0.25)
        .. controls (-2.25,0.15) and (-1.3,1.9) .. (-0.8,1.9)
        .. controls (-0.3,1.9) and (0.75,0.3) .. (0.42,-0.15)
        .. controls (0.09,-0.6) and (-1.65,-0.65) .. (-1.95,-0.25) -- cycle;

    \draw[black] (-2.05,0.15)
        .. controls (-1.75,0.55) and (-0.2,0.55) .. (0.35,0.15)
        .. controls (0.65,-0.15) and (-0.25,-1.85) .. (-0.85,-1.9)
        .. controls (-1.45,-1.95) and (-2.35,-0.28) .. (-2.05,0.15) -- cycle;

    \draw[black] (-1.2,-1.65)
        .. controls (-1.7,-1.1) and (-0.55,0.4) .. (0,0.4)
        .. controls (0.55,0.4) and (1.7,-1.1) .. (1.2,-1.65)
        .. controls (0.7,-2.2) and (-0.7,-2.2) .. (-1.2,-1.65) -- cycle;

    \draw[black] (-0.4,-0.2)
        .. controls (-0.65,0.2) and (0.25,1.92) .. (0.8,1.92)
        .. controls (1.35,1.92) and (2.15,0.55) .. (2.05,0)
        .. controls (1.95,-0.55) and (-0.15,-0.6) .. (-0.4,-0.2) -- cycle;

    \node[vertex, fill=red, text=white] at (A) {$a$};
    \node[vertex, fill=red, text=white] at (B) {$b$};
    \node[vertex, fill=red, text=white] at (C) {$c$};
    \node[vertex, fill=red, text=white] at (D) {$d$};

    \node[vertex, fill=blue, text=white] at (U) {$u$};
    \node[vertex, fill=blue, text=white] at (V) {$v$};
    \node[vertex, fill=blue, text=white] at (W) {$w$};

    \node[fill=white, inner sep=1.5pt] at (-1.2,1.95) {$h_1$};

    \node[left, fill=white, inner sep=1.5pt] at (-1.95,-0.95) {$h_2$};

    \node[below, fill=white, inner sep=1.5pt] at (0,-2.1) {$h_3$};

    \node[fill=white, inner sep=1.5pt] at (1.2,1.95) {$h_4$};
\end{scope}
\end{tikzpicture}
}
  }
  \caption{Reduction from \SPECC{} to a 3-uniform bipartite hypergraph vertex cover problem. For example, $u$ is a node at the intersection of two different-colored edges $a$ and $b$ in the \SPECC{} instance on the left. In order to resolve conflicts, at least one of $\{u,a,b\}$ must be deleted, which is equivalent covering at least one node in the edge $\{u,a,b\}$ on the right.}
  \Description{Insert}
  \label{fig:hvcreduction}
\end{figure}

\subsection{Approximations via simple reductions}
\label{sec:simplereductions}
We begin with approximation algorithms that can be obtained for any fixed $\gamma$, via simple reduction to existing problems. 
While these have some barriers in terms of scalability and implementability, they serve as a useful starting point from which to develop more direct, practical, and scalable methods. 

\paragraph{Reduction to hypergraph vertex cover.}
Consider an \SPECC{} instance $H = (V, E, C, \ell)$ with weight function $w \colon V \cup E \rightarrow \mathbbm{R}^+$. We reduce this to a vertex cover problem on a new hypergraph $\mathcal{H} = (\mathcal{V}, \mathcal{E})$ constructed as follows: (1) Treat all nodes and edges in $H$ as nodes in $\mathcal{H}$ (i.e., $\mathcal{V} = V \cup E$). (2) For $v \in V$, let the cost of covering $v$ in $\mathcal{H}$ be $\gamma w(v)$, and the cost of covering $e \in E$ be $w(e)$. Finally, (3) introduce an edge to $\mathcal{E}$ for every bad triplet $\{v, e_1, e_2\}$ in the \SPECC{} instance (i.e., $v \in e_1 \cap e_2$ and $\ell(e_1) \neq \ell(e_2)$). This construction is illustrated in Figure~\ref{fig:hvcreduction}. This is a slight generalization of the known reduction from \minecc{} to standard vertex cover~\cite{veldt2023optimal}, and immediately implies a 3-approximation algorithm for \SPECC{} via known techniques for hypergraph vertex cover~\cite{BarYehuda1985}.

The 3-uniform hypergraph resulting from this reduction has a special bipartite structure: all of its nodes can be separated into a blue set of nodes (representing nodes in the original \SPECC{} instance), and red nodes (edges in the \SPECC{} instance). Every hyperedge contains exactly two red nodes and one blue node. For this type of hypergraph vertex cover problem, we can obtain an improved 2-approximation guarantee. This applies to not only \SPECC{}, but to any 3-uniform hypergraph vertex cover problem with this structure. Let $\mathcal{H} = (\mathcal{V}, \mathcal{E}$) now be any hypergraph  where nodes can be separated into a (disjoint) red and blue set $\mathcal{V} = \mathcal{V}_r \cup \mathcal{V}_b$ and every edge $e \in \mathcal{E}$ satisfies $|e \cap \mathcal{V}_r| = 2$, and $|e \cap \mathcal{V}_b| = 1$. Let $w_v \geq 0$ denote the cost to cover $v \in \mathcal{V}$. Consider the standard LP relaxation for hypergraph vertex cover:
\begin{align}
	\label{eq:hvclp}
\begin{aligned}
	\text{min} \quad& \textstyle \sum_{v \in \mathcal{V}} w_v x_v \\
	\text{s.t.} \quad& \forall e \in \mathcal{E}:\\
	& \forall v \in \mathcal{V}
\end{aligned}
\begin{aligned}
	&\\
	&\textstyle \sum_{v \in e} x_v \geq 1 \\
	&x_v \geq 0.
\end{aligned}
\end{align}
\begin{theorem}
\label{thm:conflict-2-approx}
Let $\{x_v^*\}_{v \in \mathcal{V}}$ represent an optimal set of variables for LP~\eqref{eq:hvclp}. For $\rho \in (0,1)$, define $D_{R,\rho} = \{v \in \mathcal{V}_r \colon x_v^* \geq \tfrac{\rho}{2}\}$ and $D_{B,\rho} = \{v \in \mathcal{V}_b \colon x_v^* \geq 1-\rho\}$. Then $D_{R,\rho} \cup D_{B,\rho}$ is a feasible solution for vertex cover on $\mathcal{H}$ for every $\rho \in (0,1)$, and is a 2-approximation in expectation if $\rho$ is chosen uniformly at random from $(0,1)$.
\end{theorem}

\begin{proof}
Suppose that an edge defined on blue node $b \in \mathcal{V}_b$ and two red nodes ${r_1}, {r_2} \in \mathcal{V}_r$ is not covered by this algorithm. This would contradict LP feasibility, since this would imply that
\begin{equation*}
    x_{b} + x_{r_1} + x_{r_2} < 1-\rho + \rho/2 + \rho/2  = 1.
\end{equation*}
Observe next that if $\rho \sim \text{Unif}(0,1)$, then for every $v_r \in \mathcal{V}_r$ we have $P\left(v \in D_{R,\rho} \right) \leq 2 x_{v}^*$ and for every $v \in \mathcal{V}_b$ we have $P\left(v \in D_{B,\rho} \right) = x^*_v$. Thus, the expected cost is at most 2 times  the optimal LP value:
\begin{align*}
    E[\text{cost}]\leq \sum_{v \in \mathcal{V}_r} w_v (2x_v^*) + \sum_{v \in \mathcal{V}_b}  w_v (x_b^*) \le 2 \Big( \sum_{v \in \mathcal{V}} x_v^*\Big).
\end{align*}
\end{proof}
The proof of this result is a generalization of a recent result by Adriaens and Tatti~\cite{adriaens2026btt}. These authors proved a similar result for a problem called bad triangle traversal, which can also be viewed as a special case of bipartite 3-uniform hypergraph vertex cover.

\paragraph{Reduction to node-weighted multiway cut.} 
An instance of node-weighted multiway cut (NWMC) is given by a graph $G = (\mathcal{V}, \mathcal{E}, T)$ with $k$ terminal nodes $T = \{t_1, t_2, \hdots, t_k\} \subseteq \mathcal{V}$, and a weight $w_v \geq 0$ for each $v \in \mathcal{V}\setminus T$. The goal is to delete a minimum weight set of nodes to separate all terminals. Garg et al.~\cite{garg2004multiway} gave a $(2-2/k)$-approximation by rounding the following LP relaxation, where $\mathcal{P}$ is the set of all simple paths between pairs of terminals:
\begin{align}
	\label{eq:nwmclp}
\begin{aligned}
	\text{min} \quad& \textstyle \sum_{v \in \mathcal{V}-T} w_v x_v \\
	\text{s.t.} \quad& \forall p \in \mathcal{P}:\\
	& \forall v \in \mathcal{V}
\end{aligned}
\begin{aligned}
	&\\
	&\textstyle \sum_{v \in p} x_v \geq 1 \\
	&x_v \geq 0.
\end{aligned}
\end{align}
An instance of \SPECC{} $H = (V,E,C,\ell)$ with node weight function $w$ can be reduced to a NWMC instance $G$ as follows: 
\begin{enumerate}
    \item Treat all nodes and edges in $H$ as non-terminal nodes in $G = (\mathcal{V}, \mathcal{E}, T)$ (i.e., $\mathcal{V} = V \cup E$). For $v \in V$, assign $v$ the weight $\gamma w(v)$ and edge $e \in E$ the weight $w(e)$ in $G$. 
    \item Introduce $k$ terminal nodes $T = \{t_1, t_2, \hdots, t_k\}$ into $G$, one for each color in the \SPECC{} instance.
    \item For $v \in V$ and $e \in E$ such that $v \in e$ in the hypergraph $H$, add an edge $(v,e)$ in $G$. 
    \item For each $e \in E$ in $H$, add an edge between $e$ and $T_{\ell(e)}$ in the NWMC instance $G$. 
\end{enumerate}
This is an approximation preserving reduction, which immediately implies a $(2-2/k)$-approximation for \SPECC{}. As an interesting observation, the LP relaxations for \SPECC{} obtained by reduction to hypergraph vertex cover or by reduction to NWMC are equivalent. We include a short proof below.
\begin{observation}
\label{obs:same}
    The LP relaxation for \SPECC{} obtained by reducing to hypergraph vertex cover and using LP~\eqref{eq:hvclp} is equivalent to the LP relaxation obtained by reducing to NWMC and using LP~\eqref{eq:nwmclp}.
\end{observation}
\begin{proof}
    Consider any simple path between two terminal nodes $T_i$ and $T_j$ in the reduction to NWMC. This must always pass from $T_i$ to a node $e$ (corresponding to an edge of color $i$ in \SPECC{} instance $H$), then to a node $v$ (corresponding to a node in $e$ in $H$), then to a node $f$ in $G$ (corresponding to an edge $f$ in $H$ of color $j$), and then to $T_j$. The constraint in LP~\eqref{eq:nwmclp} associated with this path is then $x_v + x_e + x_f \geq 1$, which is exactly the 3-variable constraint associated with a bad triplet $\{v,e,f\}$ in the reduction to 3-uniform hypergraph vertex cover problem. It is not hard to see this reduction goes both directions; thus the resulting LP relaxations are equivalent.
\end{proof}

\subsection{Rounding a tighter LP relaxation}
The hypergraph vertex cover LP relaxation in~\eqref{eq:hvclp} (which is equivalent to the LP obtained via reduction to NWMC) has polynomial size in terms of the \SPECC{} instance, and as we have seen comes with strong approximation guarantees. However, explicitly constructing the reduced instance of vertex cover and explicitly forming this LP is impractical. In particular, recent work has shown that reducing standard \minecc{} to (standard, 2-uniform) vertex cover leads to extremely dense instance of vertex cover~\cite{han2026improved}. This leads to memory issues in practice even when simply trying to form the constraint matrix of the resulting LP relaxation. The reduction from \SPECC{} to (hypergraph) vertex cover leads to even more hyperedges and constraints, since it involves a constraint for every bad triplet $\{v, e, f\}$, rather than just every bad edge pair $\{e,f\}$. 

In order to avoid this computational challenge, we introduce a more direct linear programming relaxation for \SPECC{} that does not include a constraint for every bad triplet:
\begin{align}
	\label{eq:specclp}
	\begin{aligned}
		\text{min} \quad& \textstyle  \sum_{e \in E} w(e) x_e + \gamma \sum_{v \in V} w(v) y_v \\
		\text{s.t.} \quad& \forall v \in V: \quad \textstyle \sum_{c=1}^k x_v^c =(k - 1) + y_v \\
		&\forall e \in E: \quad x_v^{\ell(e)} \leq x_e + y_v \quad \forall  v \in e \\
		&x_v^c, x_e, y_v \in [0, 1] \quad \forall  c \in [k], v \in V, e \in E.
    \end{aligned}
\end{align}
Variable $x_v^c$ can be viewed as the distance between node $v$ and color $c$. If variables were restricted to be binary, $y_v \in \{0,1\}$ and $x_e \in \{0,1\}$ would be indicator variables for whether to delete node $v$ and edge $e$ respectively. The first constraint encodes the fact that each node should either be deleted, or have distance 1 from at least $k-1$ colors. The second constraint indicates that if $v \in e$, then either $v$ needs to be given color $\ell(e)$, or $v$ should be deleted, or $e$ should be deleted.

In addition to having fewer constraints in practice,
we can show that this new relaxation provides a tighter lower bound for \SPECC{}.
\begin{theorem}
\label{lem:tight}
    Let $\text{OPT}_{\text{LP}}$ be optimal solution value for the LP relaxation in~\eqref{eq:specclp}, and $\text{OPT}_\text{VC}$ be the optimal solution to the LP relaxation obtained by reducing to vertex cover and using LP~\eqref{eq:hvclp}. Then $\text{OPT}_\text{VC} \leq \text{OPT}_{\text{LP}}$.
\end{theorem}
\begin{proof}
The LP relaxation obtained via vertex cover reduction is
\begin{align}
	\label{eq:vcspecclp}
\begin{aligned}
	\text{min} \quad& \textstyle  \sum_{e \in E} w(e) x_e + \gamma \sum_{v \in V} w(v) y_v \\
	\text{s.t.} \quad& x_e + x_f + x_v \geq 1 \quad \text{ for every bad triplet } \{v, e, f\}\\
	&y_v, x_e \geq 0 \hspace{35pt} \text{ for every } v \in V, e \in E.
\end{aligned}
\end{align}
The objective function coincides with the objective for LP~\eqref{eq:specclp}, so it suffices to show that a feasible solution for LP~\eqref{eq:specclp} can be converted into a feasible solution for LP~\eqref{eq:vcspecclp} with the same objective score. This will show that the optimal score for LP~\eqref{eq:vcspecclp} cannot be larger than the optimal score for LP~\eqref{eq:specclp}.

Let $\{x_e, y_v, x_v^c \colon v \in V, e\in E, c \in[k]\}$ encode a feasible solution for LP~\eqref{eq:specclp}. Let $\{v, e, f\}$ be an arbitrary bad triplet and let $a = \ell(e)$ and $b = \ell(f)$. By the first LP constraint we know that:
\begin{align*}
    (k-1) + y_v = \sum_{i =1}^k x_v^i \leq (k-2) + x_v^a + x_v^b \implies 1 \leq x_v^a + x_v^b - y_v.
\end{align*}
We use the second constraint of LP~\eqref{eq:specclp} to see that:
\begin{align*}
    x_e + x_f + y_v \geq (x_v^a - x_v) + (x_v^b - x_v) + x_v =  x_v^a + x_v^b - y_v \geq 1.
\end{align*}
So this is feasible for LP~\eqref{eq:vcspecclp}.
\end{proof}
The proof above is a slight generalization of a previous result showing that the standard \minecc{} LP relaxation is tighter than the relaxation obtained by reducing standard \minecc{} to NWMC~\cite{veldt2023optimal}. It was also shown that the LP lower bound from the canonical \minecc{} relaxation is strictly tighter for some hypergraphs than the lower bound obtained by the NWMC relaxation~\cite{veldt2023optimal}. Thus, we know that there exist special cases (e.g., when $\gamma = \infty$) where our LP~\eqref{eq:specclp} provides a better lower bound for \SPECC{} than the lower bound obtained via the NWMC LP relaxation.

Algorithm~\ref{alg:direct-lp-rounding} can be viewed as a warm-up. It gives a simple strategy for rounding LP~\eqref{eq:specclp} into an expected 2-approximate solution.
\begin{algorithm}[t]
\caption{Simple Rounding for LP~\eqref{eq:specclp}}
\label{alg:direct-lp-rounding}
\begin{algorithmic}[1]
\Require \SPECC{} instance $H = (V, E, C, \ell)$, parameter $\gamma$ 
\Ensure Deletion sets $D_E \subseteq E$, $D_V \subseteq V$, and map $\lambda \colon V \setminus D_V \to C$
\State Solve LP~\eqref{eq:specclp} to obtain optimal values $\{x_e^*$, $y_v^*$,$x_v^{i\,*}\}$
\State Draw $\rho \in [0, \tfrac{1}{2})$ uniformly at random
\State $D_E = \{e \in E \colon x_e^* \geq \rho\}$ \Comment{Edge deletion}
\State $D_V = \{v \in V \colon y_v^* \geq 1/2 - \rho \}$ \Comment{Node deletion}
\State For every $v \in V \setminus D_V$, set $\lambda(v) = \ell(e)$ if there is some $e \in E \setminus D_E$ with $v \in e$, else set $\lambda(v)$ to an arbitrary color
\State \Return $(D_E, D_V, \lambda)$
\end{algorithmic}
\end{algorithm}
\begin{theorem}
\label{thm:direct-2-approx}
Algorithm~\ref{alg:direct-lp-rounding} is a randomized $2$-approximation for \SPECC{}.
\end{theorem}
\begin{proof}
We first prove feasibility, which is equivalent to showing that every surviving node is contained in edges of at most one color. Fix a surviving edge $e \in E \setminus D_E$ with color $c = \ell(e)$ and a surviving vertex $v \in e \setminus D_V$. By the deletion rules,
$x_e^* < \rho$ and $y_v^* < \tfrac{1}{2} - \rho$. Combined with the LP constraints this gives
\[
     x_v^{c\,*} \leq  x_e^* + y_v^* < \tfrac12.
\]
If $v$ were also contained in some surviving edge $f \in E \setminus D_E$ with color $d \neq c$, then using the same steps we could show $x_v^{d\,*} < \tfrac{1}{2}$. However, this would contradict LP constraints since:
\begin{equation*}
    (k-1) + y_v = \sum_{i = 1}^k x_v^{i,*} \leq (k-2) + \tfrac{1}{2} +x_v^{d,*} +x_v^{c,*} \implies 1 \leq x_v^{d,*} +x_v^{c,*}.
\end{equation*}
Now we prove our approximation guarantee. Since $\rho \sim \text{Unif}[0,\tfrac12)$ and $x_e^* \in [0, 1]$, the probability that edge $e$ is deleted is
\[
    \Pr[e \in D_E]=\ \Pr[\rho \leq x_e^*] \leq \frac{x_e^*}{1/2} = 2 x_e^*.
\]
Similarly, the probability that vertex $v$ is deleted is
\[
    \Pr[v \in D_V] =\Pr[\rho \geq \tfrac{1}{2} - y_v^*]\leq \frac{y_v^*}{1/2} = 2 y_v^*.
\]
The
total cost is exactly $\sum_{e \in D_E} w_e + \sum_{v \in D_V} w_v$.
By linearity of expectation, we have the 2-approximation guarantee:
\begin{align*}
    \mathbb{E}[\mathrm{cost}] &= \sum_{e \in E} w_e \Pr[e \in D_E]
        + \sum_{v \in V} w_v \Pr[v \in D_V] \leq 2 \Big( \sum_{e \in E} w_e x_e^*
                 + \sum_{v \in V} w_v y_v^* \Big)
= 2 \cdot \mathrm{OPT}_{\mathrm{LP}}.
\end{align*}
\end{proof}

Algorithm~\ref{alg:better-direct-lp-rounding} is more involved (both in terms of the algorithmic procedure and its proof) but has a better guarantee. It is closely related to the $(2-2/k)$-approximate rounding algorithm for NWMC of Chekuri and Madan~\cite{chekuri2016simple}; we have adapted their techniques to apply directly to our tighter LP relaxation. In Algorithm~\ref{alg:better-direct-lp-rounding}, $E_j$ denotes the set of hyperedges with color $j$.
\begin{algorithm}[t]
\caption{$(2-2/k)$ Rounding for LP~\eqref{eq:specclp}}
\label{alg:better-direct-lp-rounding}
\begin{algorithmic}[1]
\Require \SPECC{} instance $H = (V, E, C, \ell)$, parameter $\gamma$
\Ensure Deletion sets $D_E \subseteq E$, $D_V \subseteq V$, and map $\lambda \colon V \setminus D_V \to C$
\State Solve LP~\eqref{eq:specclp} to obtain optimal values $\{x_e^*$, $y_v^*$,$x_v^{i\,*}\}$
\State Draw $\rho \in [0, \tfrac{1}{2})$ uniformly at random
\State Draw $j$ uniformly at random from $[k]$
\State $D_E = \{e \notin E_j \colon x_e^* \geq \rho\} \cup \{e \in E_j \colon \exists v \in e,  c \neq j \text{ s.t. } x_v^{c*} < \rho \}$ 
\State $D_V = \{v \in V \colon \exists c \neq j \text{ s.t. } x_v^{c*} - y_v^* < \rho \leq x_v^{c*}\}$ 
\State For every $v \in V \setminus D_V$, set $\lambda(v) = \ell(e)$ if there is some $e \in E \setminus D_E$ with $v \in e$, else set $\lambda(v)$ to an arbitrary color
\State \Return $(D_E, D_V, \lambda)$
\end{algorithmic}
\end{algorithm}

Before proving the guarantee for Algorithm~\ref{alg:better-direct-lp-rounding}, we collect three simple facts about feasible solutions to LP~\eqref{eq:specclp}. We omit their proofs, since each follows directly from the LP constraints.
\begin{observation}
\label{obs:simple}
For all feasible solutions to LP~\eqref{eq:specclp}
    \begin{enumerate}
  \item[(a)] For every $v \in V$ and colors $c \neq c'$,\;
    $x_v^c + x_v^{c'} \ge 1 + y_v$.
  \item[(b)] For every $v \in V$ and color $c$,\; $x_v^c \ge y_v$.
  \item[(c)] For every $e \in E$, every $u \in e$ and every color $c \neq \ell(e)$,\;
    $x_u^c \ge 1 - x_e$.
\end{enumerate}
\end{observation}
In what follows, let $\{x_e, y_v, x_v^c\}$ denote an optimal solution to LP~\eqref{eq:specclp}, and recall that a \emph{bad triplet} is a set $\{v, e, f\}$ with $v \in e \cap f$ and $\ell(e) \neq \ell(f)$. We first show that Algorithm~\ref{alg:better-direct-lp-rounding} always returns a feasible solution.
\begin{lemma}
\label{lem:feasible}
For every $\rho \in (0, \frac12)$ and every $j \in [k]$, every bad triplet $\{v, e, f\}$ has at least one element in $D_V \cup D_E$. 
\end{lemma}

\begin{proof}
Let $\{v, e, f\}$ be a bad triplet and let $a = \ell(e)$ and $b = \ell(f) \neq a$. At most one of $a, b$ equals $j$, so assume without loss of generality that $a \neq j$. 
Suppose $e \notin D_E$ and $v \notin D_V$. We will show $f \in D_E$, which will complete the proof.
 
Since $\ell(e) = a \neq j$ and $e \notin D_E$, we have $x_e < \rho$. The LP constraint
for $e$ and $v$ gives $x_v^a - y_v \leq x_e < \rho$. Since $a \neq j$ and $v \notin D_V$, the condition
$x_v^a - y_v < \rho \le x_v^a$ must fail, and therefore $x_v^a < \rho < \tfrac12$.
 
\emph{Case $\ell(f) = b \neq j$.} If we assume that $f \notin D_E$, we could use the same logic as above to show that this implies $x_v^b < \rho < \tfrac12$. However, this would be a contradiction since $x_v^i < \tfrac12$ for at most one color $i$, by Observation~\ref{obs:simple}(a). Hence, $f \in D_E$.
 
\emph{Case $\ell(f) = b = j$.} Node $v \in f$ and color $a \neq j$ satisfy $x_v^a < \rho$,
so by the definition of $D_E$ we have $f \in D_E$.
 
\end{proof}

Next, we bound the probability that each node and each hyperedge is deleted.
\begin{lemma}
\label{lem:node}
For every node $v \in V$,\;
$\Pr[v \in D_V] \le \bigl(2 - \frac{2}{k}\bigr) y_v$.
\end{lemma}
\begin{proof}
For each color $c$ let $a_c = x_v^c - y_v$ and
$I_c = (a_c,\, x_v^c] \cap (0, \frac12)$. By Observation~\ref{obs:simple}(b),
$a_c \ge 0$. Each $I_c$ has length at most $x_v^c - a_c = y_v$, and
$v \in D_V$ if and only if $\rho \in \bigcup_{c \neq j} I_c$.
Let $N = \{c \in [k] : I_c \neq \emptyset\}$.
 
\emph{Case $|N| = 0$.} Then $\Pr[v \in D_V] = 0$.
 
\emph{Case $|N| = 1$.} Let $N = \{c_1\}$. Then $v \in D_V$ only if $j \neq c_1$ and
$\rho \in I_{c_1}$, so
\[
  \Pr[v \in D_V] = \Bigl(1 - \frac1k\Bigr) \cdot 2\,|I_{c_1}|
  \le \Bigl(2 - \frac2k\Bigr) y_v .
\]
 
\emph{Case $|N| = t \ge 2$.} If $c \in N$, this means $(a_c, x_v^c]$ has nonempty overlap with $(0,\tfrac12)$ and hence $a_c < \frac12$. For $c \in N$ pick any
$c' \in N$ with $c' \neq c$. By Observation~\ref{obs:simple}(a) we have
\[
  x_v^c \ge 1 + y_v - x_v^{c'} = 1 - a_{c'} > \tfrac12 ,
\]
so $I_c = (a_c, \frac12)$ for every $c \in N$. Order $N$ as $c_1, c_2, \dots, c_t$ with
$a_{c_1} \le a_{c_2} \le \cdots \leq a_{c_t}$. The intervals are nested,
$I_{c_1} \supseteq I_{c_2} \supseteq \cdots \supseteq I_{c_t}$, so
$\bigcup_{c \neq j} I_c$ equals $(a_{c_1}, \frac12)$ if $j \neq c_1$ and
$(a_{c_2}, \frac12)$ if $j = c_1$. Hence
\begin{align*}
  \Pr[v \in D_V]
  &= \Bigl(1 - \frac1k\Bigr) \cdot 2\Bigl(\frac12 - a_{c_1}\Bigr)
    + \frac1k \cdot 2\Bigl(\frac12 - a_{c_2}\Bigr)\\
  &= \frac{2}{k}\Bigl[(k-1)\Bigl(\frac12 - a_{c_1}\Bigr) + \Bigl(\frac12 - a_{c_2}\Bigr)\Bigr] \\
  &\leq 2 \frac{(k-1)}{k} \left(\frac12 - a_{c_1} + \frac{1}{2} - a_{c_2}\right) \\
  &= 2\Big(1 - \frac1k\Big) \left(1 - a_{c_1} - a_{c_2} \right)
\end{align*}
where we have use the fact that $k - 1 \ge 1$ and $\frac12 - a_{c_2} > 0$. By Observation~\ref{obs:simple}(a) applied to $c_1, c_2$,
\[
  a_{c_1} + a_{c_2}  = x_v^{c_1} + x_v^{c_2} - 2y_v \ge 1 - y_v \implies 1 - a_{c_1} - a_{c_2} \le y_v.
\]
Therefore
$\Pr[v \in D_V] \le \bigl(2 - \frac2k\bigr) y_v$.
\end{proof}
 
\begin{lemma}
\label{lem:edge}
For every hyperedge $e \in E$,\;
$\Pr[e \in D_E] \le \bigl(2 - \frac{2}{k}\bigr) x_e$.
\end{lemma}
 
\begin{proof}
Let $z_e = \min\{x_u^c : u \in e,\ c \neq \ell(e)\}$.
Observation~\ref{obs:simple}(c) implies that $z_e \ge 1 - x_e$. If $\ell(e) \neq j$, then
$e \in D_E$ if and only if $\rho \le x_e$. If $\ell(e) = j$, then
$e \in D_E$ if and only if $\rho > z_e$. Since $\Pr[\ell(e) = j] = \frac1k$
and $j$ is independent of $\rho$,
\[
  \Pr[e \in D_E]
  = \Bigl(1 - \frac1k\Bigr) \Pr[\rho \le x_e] + \frac1k \Pr[\rho > z_e] .
\]
Here $\Pr[\rho \le x_e] = \min\{1, 2x_e\}$ and
$\Pr[\rho > z_e] = 2\max\{0, \frac12 - z_e\} \le 2\max\{0, x_e - \frac12\}$. If $x_e \le \frac12$, the second term is $0$, so $\Pr[e \in D_E] = \bigl(1 - \frac1k\bigr) 2 x_e$. Assume then that $x_e > \tfrac12$. Then since $2x_e > 1$ we have
\begin{align*}
    \Pr[e \in D_E] &\le \Big(1 - \frac1k\Big) + \frac{2x_e - 1}{k} = 1 - \frac2k + \frac{2x_e}{k} \\
    & \le \Big(1 - \frac2k\Big) 2x_e + \frac{2x_e}{k} = 2\Big(1 - \frac1k\Big) x_e.
\end{align*}
\end{proof}

\begin{theorem}
\label{thm:better}
Algorithm~\ref{alg:better-direct-lp-rounding} is a $(2-2/k)$-approximation for \SPECC{}.
\end{theorem}
\begin{proof}
Lemma~\ref{lem:feasible} establishes feasibility. The cost of the output is exactly $\sum_{v \in D_V} \gamma w(v) + \sum_{e \in D_E} w(e)$. Lemmas~\ref{lem:node} and~\ref{lem:edge} show the expected cost is within a factor $2-\tfrac2k$ of the LP lower bound:
\begin{align*}
  \mathbb{E}[\text{cost}]
  &= \sum_{v \in V} \gamma w(v) \Pr[v \in D_V] + \sum_{e \in E} w(e) \Pr[e \in D_E] \\
  &\le \Bigl(2 - \frac2k\Bigr)\Bigl(\sum_{v \in V} \gamma w(v) y_v
      + \sum_{e \in E} w(e) x_e\Bigr).
\end{align*}
The right-hand side is $(2-2/k)$ times the optimal LP value, which is at most the optimal \SPECC{} cost.
\end{proof}

Both algorithms can be derandomized by trying all possible $\rho$ parameters (and for Algorithm~\ref{alg:better-direct-lp-rounding}, all possible choices of $j$) that lead to distinct deletion sets, and taking the best result.

\section{Fast algorithms with nested deletions}
\label{sec:nested}

The algorithms in the previous section all run in polynomial time and work for any choice of global node deletion parameter $\gamma \geq 0$. However, they still face two practical challenges. 

\paragraph{Challenge 1: scalability.} As discussed, explicitly forming the hypergraph vertex cover problem (or writing down the resulting LP relaxation) is prohibitively expensive. Our direct LP rounding algorithms (Algorithms~\ref{alg:direct-lp-rounding} and~\ref{alg:better-direct-lp-rounding}) partially mitigate this, but still rely on black-box LP solvers and have high theoretical runtimes. 

\paragraph{Challenge 2: unrelated solutions when varying $\gamma$.}
In practice, there is unlikely to be a single ``correct'' choice of parameter $\gamma$ for determining which nodes should be deleted or viewed as outliers. Rather, a natural approach is to find approximatie solutions for a range of $\gamma$ values and observe different points at which different nodes are deleted. One downside of the algorithms in Section~\ref{sec:specc} is that the solutions they output for different choices of $\gamma$ are not necessarily correlated in any way. In particular, these algorithms may delete a node $v$ for some large value of $\gamma$, but then not delete $v$ when considering a smaller penalty for node deletion $\gamma' < \gamma$. 

Motivated by Challenge 2, we introduce a new notion of a 
\emph{nested} approximation algorithm.
\begin{property}
\label{prop:nested}
    A \SPECC{} algorithm is a \textbf{nested $p$-approximation} algorithm if, for any sequence of parameters $0\leq \gamma_1 < \gamma_2 < \cdots < \gamma_t$, it returns a family of deletion sets $\{(D_E^{(i)}, D_V^{(i)}) \colon i \in [t] \}$ such that for every $i \in [t]$, $(D_E^{(i)}, D_V^{(i)})$ is a $p$-approximation for \SPECC{}($\gamma_i$), and for every $i \in [t-1]$ we have $D_V^{(i+1)} \subseteq D_V^{(i)}$ and $D_E^{(i)} \subseteq D_E^{(i+1)}$.
\end{property}
In other words, as $\gamma$ (the node-deletion penalty) increases, the set of deleted nodes is monotonically shrinking and the set of deleted edges is monotonically growing. Although this property is intuitive, we shall see in Section~\ref{sec:experiments} that is it not always satisfied by our LP-based algorithms. It is also not obvious a priori whether such an algorithm is guaranteed to exist for different values of $p$, even if we allow unlimited compute time. However, we will show how to design an algorithm that is a nested 3-approximation. Furthermore, this algorithm is purely combinatorial and has a runtime that is linear in terms of the hypergraph size for each value of $\gamma$ considered, addressing Challenge 1. 
To obtain this algorithm, we first design an algorithm for a sequence of more general vertex cover (VC) problems over a class of 3-uniform bipartite hypergraphs, which may be of independent interest.

\newcommand{\bhvc}{\textsc{B3UHVC}}
\subsection{Nested vertex cover in bipartite hypergraphs}
Let $\mathcal H=(\mathcal{V}_r \cup \mathcal{V}_b,\mathcal E)$ be a 3-uniform bipartite hypergraph with disjoint red and blue node sets ($\mathcal{V}_r$, $\mathcal{V}_b$), where every $e \in \mathcal{E}$ satisfies $|e \cap \mathcal{V}_r| = 2$ and $|e \cap \mathcal{V}_b| =1 \neq \varnothing$.
Let $w_v \geq 0$ denote a fixed weight for $v \in \mathcal{V}_r \cup \mathcal{V}_b$ and $\gamma \in[0,\infty)$ be a scaling parameter for blue nodes.
Let $\bhvc(\mathcal{H},\gamma)$ denote the Bipartite $3$-uniform Hypergraph Vertex Cover problem defined over $\mathcal{H}$ in which the cost for deleting $v \in \mathcal{V}_r$ is $c_\gamma(v) = w_v$ and the cost for deleting $v \in \mathcal{V}_b$ is $c_\gamma(v) = \gamma w_v$. Consider a non-decreasing collection of parameters
$\Gamma= \langle \gamma_1,\ldots,\gamma_t \rangle$, and let $\text{OPT}_i$ be the optimal vertex cover solution value for $\bhvc(\mathcal{H},\gamma_i)$.
Our goal is to find a hypergraph vertex cover $C_i$ for each $\gamma_i$, such that $\text{cost}(C_i) \coloneqq \sum_{v \in C_i} c_{\gamma_i}(v) \leq 3 \text{OPT}_i$, while ensuring a \emph{nestedness} property:
once a red node is selected to be in the cover, it remains selected; and once a blue node is not selected, it cannot be selected for a larger parameter. We prove that these goals are accomplished by Algorithm~\ref{alg:covernested_41}. In order to prove the result, recall first of all that for 
a fixed parameter $\gamma_i$, the LP relaxation for vertex cover on $\mathcal{H}$, and its dual LP, are given by:
\begin{align*}
\begin{aligned}
&\textbf{LP relaxation} \\ 
\min\quad
    &\sum_{v \in \mathcal{V}_r} w_v x_v + \sum_{v \in \mathcal{V}_b} \gamma w_v x_v\\
\text{s.t.} \quad
    &\sum_{v \in h}x_v \ge 1 \quad \forall h \in \mathcal E\\
    &x_v \ge 0 \quad \forall v \in \mathcal V  \\ \\
\end{aligned}
\hspace{50pt}
\begin{aligned}
&\textbf{Dual LP} \\
\max\quad
    &\sum_{h \in \mathcal E}y_h\\
\text{s.t.}\quad
    &\sum_{h \ni v}y_h\le w_v && \forall v \in \mathcal{V}_r\\
    &\sum_{h \ni v}y_h\le \gamma_i w_v && \forall v \in \mathcal{V}_b\\
    &y_h \ge 0 && \forall h \in \mathcal E.
\end{aligned}
\end{align*}
\begin{algorithm}[t]
\caption{\textsc{CoverNested}$(\mathcal H,\Gamma)$}
\label{alg:covernested_41}
\begin{algorithmic}[1]
\Require Bipartite 3-uniform hypergraph $\mathcal H=(\mathcal{V}_r\cup \mathcal{V}_b,\mathcal E)$ and $\Gamma=\langle\gamma_1,\ldots,\gamma_t\rangle$, where
$0\le\gamma_1\le\cdots\le\gamma_t$
\State $s_v\gets0$ for all $v\in \mathcal{V}_r \cup \mathcal{V}_b$
\State $y_h\gets0$ for all $h \in \mathcal{E}$
\State $\mathcal S\gets\langle\,\rangle$
\For{$i=1$ \textbf{to} $t$}
    \For{$h=\{r_1,r_2,b\}\in\mathcal E$ where $r_1, r_2 \in \mathcal{V}_r$ and $b \in \mathcal{V}_b$}
        \State $\delta\gets\min\{w_{r_1}-s_{r_1},w_{r_2}-s_{r_2},\gamma_iw_b-s_b\}$
        \For{each $v\in h$}
            \State $s_v\gets s_v+\delta$
        \EndFor
        \State $y_h \gets y_h + \delta$
    \EndFor
    \State $C_{i}^{r}\gets\{v\in \mathcal{V}_r:s_v=w_v\}$ \label{ln:cr}
    \State $C_{i}^{b}\gets\{v\in \mathcal{V}_b:s_v=\gamma_i w_v\}$ \label{ln:cb}
    \State $C_{i}\gets C_{i}^{r}\cup C_{i}^{b}$
    \State Append $C_{i}$ to $\mathcal S$
\EndFor
\State \Return $\mathcal S$
\end{algorithmic}
\end{algorithm}
We know from primal-dual theory that if $\{y_h^{(i)}\}$ is feasible for this fixed $\gamma_i$, then its LP objective is a lower bound for the optimal primal LP solution, which is in turn a lower bound vertex cover:
\begin{equation}
    {\sum_{h \in \mathcal{E}} y_h^{(i)} \leq \text{OPT}_i, \quad \text{ if } \{y_h^{(i)}\} \text{ is feasible for dual LP for $\gamma_i$.}}
\end{equation}

As a 3-uniform hypergraph vertex cover problem, one can obtain a 3-approximation quickly for a fixed $\gamma_i$ using the standard local-ratio algorithm~\cite{BarYehuda1985}. 
This method sets all dual variables to zero, greedily builds a maximal feasible dual solution, then covers nodes corresponding to tight constraints. Algorithm~\ref{alg:covernested_41} adapts this by applying the same mechanism multiple times, in a way that iteratively \emph{updates} previous dual variables whenever it visits a new $\gamma$ parameter, rather than restarting the local-ratio algorithm for each $\gamma$. 
\begin{theorem}
\label{thm:nestingvc}
For each $\gamma_i\in\Gamma$, the vertex cover $C_{i}$ generated
by Algorithm~\ref{alg:covernested_41} satisfies
$\operatorname{cost}(C_{i})\le3\operatorname{OPT}_{\gamma_i}$. For all $i \in [t-1]$ we have
\[
    C_{i+1} \cap \mathcal{V}_b \subseteq C_{i} \cap \mathcal{V}_b, \quad
    C_{i} \cap \mathcal{V}_r \subseteq C_{i+1} \cap \mathcal{V}_r.
\]
\end{theorem}

\begin{proof}
Algorithm~\ref{alg:covernested_41} always updates dual variables $y_h$ and node variables $s_v$ in such a way that
$s_v=\sum_{h \ni v} {y_h}$. After the $i$th iteration,  
Algorithm~\ref{alg:covernested_41} adds a node to $C_{i}$ if and only if its dual constraint is tight (lines~\ref{ln:cr} and~\ref{ln:cb}).
The vertex cover cost is therefore:
\[
\begin{aligned}
    \sum_{v \in C_{i}}c_{\gamma_i}(v)
    =\sum_{v \in C_{i}}\sum_{h\ni v}y_h =\sum_{h \in \mathcal E}|C_{i}\cap h|\,y_h \le 3 \sum_{h \in \mathcal E}y_h
      \le 3 \text{OPT}_{i}.
\end{aligned}
\]
This proves the desired 3-approximation guarantee for every $\gamma_i$.

Consider now a red node $v\in \mathcal{V}_r$, which is added to $C_{i}$ if and only if its dual constraint is tight ($\sum_{h\ni v}y_h= w_r$) after iteration $i$. The dual variables are all non-decreasing throughout Algorithm~\ref{alg:covernested_41}, and the cost of deleting $v \in \mathcal{V}_r$ is constant throughout even when $\gamma_i$ is increased. Hence, once this constraint is tight, it remains tight for all subsequent iterations. Thus, $v$ belongs to every subsequent vertex cover $C_j$ for $j > i$.

Next consider a node $b \in \mathcal{V}_b$. If $b \notin C_{i}$, then its dual constraint must not be tight after the $i$th iteration, i.e., $s_{b} <\gamma_i w_{b}$. Because $C_{i}$ is a vertex cover, every hyperedge is covered before proceeding to the next parameter $\gamma_{i+1}$. Because each hyperedge contains only one blue vertex, every hyperedge containing $b$ must contain a covered red node after iteration $i$. As we have just shown, if a red node is covered in iteration $i$, it will be covered in all future iterations. Therefore, the algorithm will never increase $y_h$ for any $h \ni b$, and $s_b$ will remain fixed. Hence, for every $j > i$ we know that $s_b < \gamma_i < \gamma_j$ which implies that $b \notin C_j$.
\end{proof}

\subsection{Nested \SPECC{} 3-approximation}
As noted in Section~\ref{sec:simplereductions}, explicitly forming the bipartite vertex cover problem corresponding to \SPECC{} is impractical. However, we can effectively run Algorithm~\ref{alg:covernested_41} directly on a \SPECC{} instance using an implicit implementation. In other words, rather than applying steps to red and blue nodes in some vertex cover problem, we apply the same mechanics directly to the \SPECC{} hypergraph. This techniques builds on the work of Veldt~\cite{veldt2023optimal}, who developed a fast combinatorial $2$-approximation for standard \minecc{} by implicitly reducing to standard (2-uniform) vertex cover. We generalize this in two ways: (1) for \SPECC{} we must consider an implicit reduction to 3-uniform hypergraph vertex cover, and (2) we are solving a sequence of vertex cover problems (as $\gamma$ varies), rather than just 1.

Observe that the core step in Algorithm~\ref{alg:covernested_41} is to (repeatedly) iterate through edges in the hypergraph vertex cover problem. The key observation for our implicit implementation is that we can do so by first iterating through nodes in $V$, and then for each node we can iterate through pairs of edges $e, f$ that contain $v$. This allows us to visit all bad edge triplets $\{e,f,v\}$, which is the same as visiting hyperedges in the (implicit) reduced hypergraph vertex cover problem. Following the approach of Veldt~\cite{veldt2023optimal}, we assume that for each node $v \in V$ we have stored its hyperedge labels $L_E(v) = \begin{bmatrix} e_1, e_2, \hdots, e_{d_v} \end{bmatrix}$, ordered by colors. Initialize a pointer $f$ to the front edge $e_1$ in $L_E(v)$ and pointer $b$ to the back edge $e_{d_v}$ in $L_E(v)$. If $\ell(e_1) \neq \ell(e_{d_v})$, then $\{v, e_1, e_{d_v}\}$ is a bad triplet and we must delete either the node or one of the two edges. We do so by decreasing dual weights associated with nodes and edges, following the mechanics of Algorithm~\ref{alg:covernested_41}. We can then immediately cover (i.e., delete) the node or edge that ends up with a tight dual constraint. If the node becomes tight, then we know all other bad triplets involving the node are taken care of, so we can move on with the next node in $V$. If one of the edges $e \in \{e_1, e_{d_v}\}$ becomes tight, then all other bad triplets involving $v$ and $e$ are covered, so we can increment either the front or back pointer.

Algorithm~\ref{alg:nested-mixed} formalizes this process for a sequence of parameters $\Gamma = \langle \gamma_1, \ldots, \gamma_t \rangle$.
\begin{algorithm}[H]
\caption{Nested Combinatorial Approximation for \SPECC{}}
\label{alg:nested-mixed}
\begin{algorithmic}[1]
\Require Edge-colored hypergraph $H=(V,E,C,\ell)$; weights $w:V\cup E\to\mathbb{R}_{\ge 0}$;
thresholds $0<\gamma_1<\gamma_2<\cdots<\gamma_t$; for each $v\in V$, incident edges
$L_E(v)=[e_1,\dots,e_{d_v}]$ sorted by color
\Ensure Deletion sets $(D_V^{(j)},D_E^{(j)})$ for $j=1,\dots,t$
\State $s_x\gets 0$ for all $x\in V\cup E$ \Comment{dual load; never reset}
\State $D_E\gets\{e\in E: w(e)=0\}$
\For{$j=1,\dots,t$}
  \State $D_V^{(j)}\gets\emptyset$
  \ForAll{$v\in V$}
    \State $f\gets 1$; \ $b\gets d_v$
    \While{$f<b$ \textbf{and} $\ell(e_f)\neq\ell(e_b)$}
      \If{$s_{e_f}=w(e_f)$}
        \State $f\gets f+1$
        \State \textbf{continue} \Comment{skip tight front edge}
      \EndIf
      \If{$s_{e_b}=w(e_b)$}
        \State $b\gets b-1$
        \State \textbf{continue} \Comment{skip tight back edge}
      \EndIf
        \State $\delta\gets\min\{\gamma_j w(v)-s_v,\ w(e_f)-s_{e_f},\ w(e_b)-s_{e_b}\}$

      \State $s_v\gets s_v+\delta$; \ $s_{e_f}\gets s_{e_f}+\delta$; \ $s_{e_b}\gets s_{e_b}+\delta$
      \If{$s_{e_f}=w(e_f)$}
        \State $D_E\gets D_E\cup\{e_f\}$
      \EndIf
      \If{$s_{e_b}=w(e_b)$}
        \State $D_E\gets D_E\cup\{e_b\}$
      \EndIf
      \If{$s_v=\gamma_j w(v)$}
        \State $D_V^{(j)}\gets D_V^{(j)}\cup\{v\}$
        \State \textbf{break} \Comment{edges are recorded before this check}
      \EndIf
    \EndWhile
  \EndFor
  \State $D_E^{(j)}\gets D_E$
\EndFor
\State \Return $\{(D_V^{(j)},D_E^{(j)})\}_{j=1}^{t}$
\end{algorithmic}
\end{algorithm}
\begin{theorem}
\label{thm:nested}
Algorithm~\ref{alg:nested-mixed} is a nested 3-approximation for \SPECC{} that runs in $O(|\Gamma|\cdot \sum_{v \in V} d_v)$ time.
\end{theorem}
\begin{proof}
    The approximation guarantee immediately follows from the proof of Theorem~\ref{thm:nestingvc}, since this algorithm is an implicit implementation of Algorithm~\ref{alg:covernested_41}. The nestedness property also follows immediately from Theorem~\ref{thm:nestingvc} and Algorithm~\ref{alg:covernested_41} for the same reason. For the runtime, note that for each parameter $\gamma_i$, Algorithm~\ref{alg:nested-mixed} visits each node once. Each time it visits the node, it does at most $O(d_v)$ work, since this is the length of $L_E(v)$ and at each step we either update one of the two pointers to $L_E(v)$, or cover node $v$ and then move on.
\end{proof}

\section{Experiments}
\label{sec:experiments}

\begin{table}[t]
\centering
\setlength{\tabcolsep}{3pt}
\caption{Dataset statistics. A node is
\emph{conflicted} if its hyperedges have two or more colors.}
\label{tab:datasets}
\begin{tabular}{lrrrr}
\toprule
Dataset & Nodes & Hyperedges & Colors & Conflicted nodes \\
\midrule
Trivago        & 207{,}974 & 247{,}362 & 55 & 48{,}389 \\
MAG-10         &  80{,}198 &  51{,}889 & 10 & 14{,}846 \\
Walmart-Trips  &  88{,}837 &  65{,}898 & 44 & 46{,}753 \\
Cooking        &   6{,}714 &  39{,}774 & 20 &  4{,}117 \\
Diseases       &     132   &   4{,}961 & 41 &     51  \\
Science        &  45{,}199 &  92{,}461 & 22 &  6{,}656 \\
Big Science    & 325{,}485 & 1{,}173{,}986 & 23 & 56{,}778 \\
\bottomrule
\end{tabular}
\end{table}

We implement our algorithms and run numerical experiments on various real-world hypergraph datasets, including several where it is particularly natural to avoid outliers in clustering.
Our first goal is to observe how well our algorithms perform in terms of solution quality and runtime, especially relative to our worst-case theoretical analysis. 
Since allowing node deletions is motivated by the goal of detecting outlier nodes in categorical clustering applications, we are also particularly interested in exploring which types of nodes that are deleted across different datasets and parameter settings. In particular, we will explore how solutions change as we vary node weights $w(v)$ and the node deletion parameter $\gamma$.

\subsection{Experimental setup and datasets}
Our algorithms are implemented in Julia and run on a 2021 14-inch MacBook Pro with an Apple M1 Pro processor (8 cores: 6 performance and 2 efficiency) and 16 GB
of RAM, running macOS 14.1.1. We use Gurobi optimization software for solving LPs. For LP-based methods, we show results just for Algorithm~\ref{alg:better-direct-lp-rounding}. We implemented Algorithm~\ref{alg:direct-lp-rounding} as well; this method also performs well but we do not report results as they are comparable in speed and slightly worse in terms of approximation factor. We specifically run the deterministic version that effectively tries all possible colors $i$ and rounding threshold $\rho$. The runtime bottleneck for this algorithm is solving the LP itself, so the time spent rounding is negligible. We refer to our implementation of Algorithm~\ref{alg:nested-mixed} as \textsf{Nested}. Our code is publicly available at: \url{https://github.com/gonebanana1729/SPECC-Experiments}.

\paragraph{Datasets.} We consider datasets summarized in Table~\ref{tab:datasets}. This includes existing benchmark real-world hypergraphs~\cite{amburg2020clustering,veldt2023optimal}. As part of our contributions, we also derive new datasets which we model for the first time as edge-colored hypergraphs. By design, these capture two situations where identifying outlier nodes is particularly natural. The first is a new dataset called Stackexchange-science (abbreviated as Science). In this hypergraph, each node is an account on \url{stackexchange.com}, a hyperedge is the collection of users that interaction with a question (either to ask it, answer it, or comment on it). Hyperedge colors correspond to the sub-site the question was posted on (e.g., Mathematics, Physics, Bioacoustics, etc.). We restrict specifically to questions on 22 sites under the ``Science'' category. We built this hypergraph from the public Stack Exchange Data Explorer, queried on 9 October 2026. We build two hypergraphs: the first is a smaller hypergraph involving all questions asked in 2024 (Science). We also build a larger hypergraph (Big Science) with over 1 million edges, that includes hyperedges for all questions asked from 2018-2024. Another new ECC dataset is Diseases, derived from the SymbPredict dataset~\cite{diseases} that links symptoms to patients with various types of diseases. This data can naturally be modeled as an edge-colored hypergraph where nodes are symptoms (e.g., skin rash), hyperedges are collections of symptoms exhibited by a sick patient, and the hyperedge color is the disease the patient is diagnosed with.

We also use four ECC datasets from prior work.

\noindent\textbf{Cooking}: each node is an ingredient, each hyperedge is a recipe, and color indicates cuisine. This was derived from the What's Cooking dataset on Kaggle~\cite{kaggle2015cooking} and first modeled as an edge-colored hypergraph by Amburg et al.~\cite{amburg2020clustering}.

\noindent\textbf{Walmart-Trips}: each node is a Walmart product, hyperedges are shopping baskets (i.e., collections of products purchased on a single shopping trip), and color indicates product type. This was derived from a Walmart-Trips Kaggle dataset~\cite{kaggle2015walmart} and first modeled as an edge-colored hypergraph by Amburg et al.~\cite{amburg2020clustering}.

\noindent\textbf{MAG-10}: nodes are authors, hyperedges are papers published by a team of authors, and colors indicate paper conference venue. The original data was obtained from the Microsoft Academic Graph and modeled as an edge-colored hypergraph by Amburg et al.~\cite{amburg2020clustering}.

\noindent\textbf{Trivago}: nodes are travel rentals from the travel website Trivago, hyperedges are rentals that are clicked-on during the same browsing session by a single user, and colors are locations where the browsing session occurred (e.g., Spain for Trivago.es or USA for Trivago.com). This was derived from the Recsys challenge 2019 data~\cite{trivagodata} and first presented as an edge-colored hypergraph in~\cite{veldt2023optimal}.

\begin{figure}[!t]
  \centering
  \subfloat[Ratio, LP rounding\label{fig:approx-lp}]
    {\includegraphics[width=.25\linewidth]{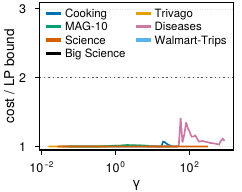}}\hfill
  \subfloat[Ratio, nested\label{fig:approx-nested}]
    {\includegraphics[width=.25\linewidth]{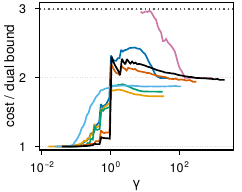}}\hfill
  \subfloat[Runtime, LP rounding\label{fig:runtime-lp}]
    {\includegraphics[width=.25\linewidth]{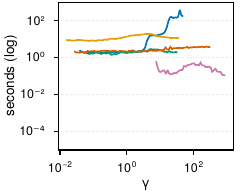}}\hfill
  \subfloat[Runtime, nested\label{fig:runtime-nested}]
    {\includegraphics[width=.25\linewidth]{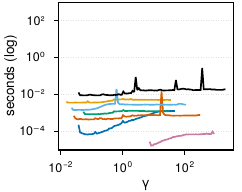}}
  \caption{LP rounding and the \textsf{Nested} algorithm as $\gamma$ varies, $w(v)=1$.
    (a) Cost divided by the LP lower bound; (b) cost divided by the nested algorithm's dual
    lower bound; dotted lines mark upper bounds on theoretical guarantees (2 and 3). (c, d) Time per $\gamma$ in seconds,
    log scale (\textsf{LP}: model setup, Gurobi solve and rounding. \textsf{Nested}: time for iteration for $\gamma_i$). Line colors are the
    same in all panels; the legend is in (a). Curves for different datasets start in different places since different datasets have different non-trivial parameter regimes, i.e., solutions that do not just delete all nodes. Section~\ref{sec:grid} explains how we chose each range.}
  \Description{Four small panels with one line per dataset. Top: approximation ratio against gamma
    for LP rounding and for the nested algorithm. Bottom: running time per gamma on a log scale.}
  \label{fig:runtimeapprox}
\end{figure}

\subsection{Approximation ratio and running time}
\label{sec:ratio-runtime}

\begin{table}[t]
  \caption{Non-nestedness of deterministic LP rounding: number of conflicted nodes that are kept at some $\gamma$
    and deleted again at a larger $\gamma$ (share of conflicted nodes). The nested algorithm has none by
    construction.}
  \label{tab:flips}
  \normalsize
  \setlength{\tabcolsep}{3pt}
  \begin{tabular}{lrrrr}
\toprule
Dataset & Conflicted & $w=1$ & $w=\sqrt{\deg}$ & $w=\deg$ \\
\midrule
Cooking & 4,117 & 33 (0.8\%) & 6 (0.1\%) & 1 (0.02\%) \\
Trivago & 48,389 & 690 (1.4\%) & 1,178 (2.4\%) & 719 (1.5\%) \\
MAG-10 & 14,846 & 268 (1.8\%) & 379 (2.6\%) & 244 (1.6\%) \\
Diseases & 51 & 14 (27.5\%) & 5 (9.8\%) & 13 (25.5\%) \\
Science & 6,656 & 22 (0.3\%) & 21 (0.3\%) & 35 (0.5\%) \\
\bottomrule
\end{tabular}
\end{table}

\begin{figure}[t]
  \centering
  \subfloat[Ratio, LP rounding\label{fig:approx-lp-sqrtdeg}]
    {\includegraphics[width=.25\linewidth]{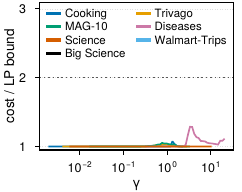}}\hfill
  \subfloat[Ratio, nested\label{fig:approx-nested-sqrtdeg}]
    {\includegraphics[width=.25\linewidth]{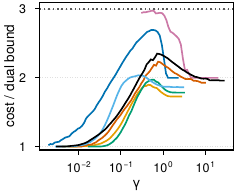}}\hfill
  \subfloat[Runtime, LP rounding\label{fig:runtime-lp-sqrtdeg}]
    {\includegraphics[width=.25\linewidth]{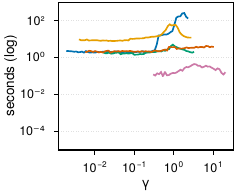}}\hfill
  \subfloat[Runtime, nested\label{fig:runtime-nested-sqrtdeg}]
    {\includegraphics[width=.25\linewidth]{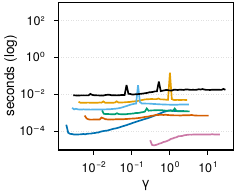}}
  \caption{Results for LP rounding and the \textsf{Nested} algorithm as $\gamma$ varies, $w(v)= \sqrt{d_v}$. See caption of Figure~\ref{fig:runtimeapprox} for more details.}
  \label{fig:runtimeapproxsqrt(deg)}
\end{figure}

\begin{figure}[t]
  \centering
  \subfloat[Ratio, LP rounding\label{fig:approx-lp-deg}]
    {\includegraphics[width=.25\linewidth]{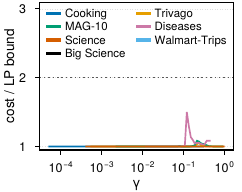}}\hfill
  \subfloat[Ratio, nested\label{fig:approx-nested-deg}]
    {\includegraphics[width=.25\linewidth]{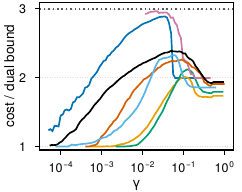}}\hfill
  \subfloat[Runtime, LP rounding\label{fig:runtime-lp-deg}]
    {\includegraphics[width=.25\linewidth]{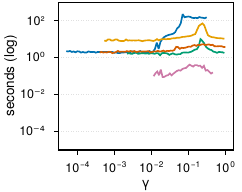}}\hfill
  \subfloat[Runtime, nested\label{fig:runtime-nested-deg}]
    {\includegraphics[width=.25\linewidth]{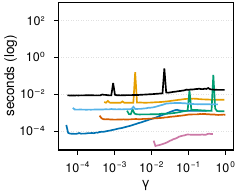}}
  \caption{LP rounding and the \textsf{Nested} algorithm as $\gamma$ varies, $w(v)=d_v$. See Figure~\ref{fig:runtimeapprox}.}
  \Description{Four small panels with one line per dataset. Top: approximation ratio against gamma
    for LP rounding and for the nested algorithm. Bottom: running time per gamma on a log scale.}
  \label{fig:runtimeapproxdeg}
\end{figure}

Figure~\ref{fig:runtimeapprox} compares the solution cost of each algorithm relative to the LP lower bound, along with their running times across different values of $\gamma$, under unit node weights. The results demonstrate that both algorithms often perform substantially better in practice than their theoretical approximation guarantees. The LP-based approach generally produces higher-quality solutions than \textsf{Nested}, with the cost-to-LP-lower-bound ratio remaining below $1.5$ in most cases. However, this improvement in solution quality comes at the expense of scalability. While the nested algorithm remains computationally efficient, on Walmart-Trips, the LP did not finish within our one-hour
per-solve limit at one value of $\gamma$, and other solves took 14 to
37 minutes. Solving it at all grid points was therefore
impractical, so we run only the nested algorithm on this dataset. We also only ran \textsf{Nested} on Big Science, as this is by far the largest hypergraph considered in our experiments. Figure~\ref{fig:runtimeapprox} uses unit node weights. Figures~\ref{fig:runtimeapproxsqrt(deg)} and~\ref{fig:runtimeapproxdeg} use using $w(v) = \sqrt{d_v}$ and $w(v) = {d_v}$, respectively. The same basic trends (LP method achieves better approximation factors, but is slower) hold in these cases as well.

Additionally, the LP-based approach does not satisfy the nestedness property (Property~\ref{prop:nested}) as $\gamma$ varies, a phenomenon particularly evident in the Diseases dataset. Table~\ref{tab:flips} quantifies violations of the nested property for the LP algorithm, counting conflicted nodes that are retained at one value of $\gamma$ but deleted at a larger value. We observe that, under unit node weights, 27.5\% of the nodes in the diseases dataset violate the nested property, highlighting the motivation for our nested algorithm. 

\paragraph{How we choose \texorpdfstring{$\gamma$}{gamma} values for grid.}
\label{sec:grid}
For small values of $\gamma$, the optimal \SPECC{} solutions (and solutions returned by approximation algorithms) may be a trivial solution where all nodes are deleted. Furthermore, after a certain point the algorithms cease deleting nodes and only delete edges. The non-trivial parameter regime varies depending on dataset size and other characteristics.
In Figure~\ref{fig:runtimeapprox},~\ref{fig:runtimeapproxdeg}, and~\ref{fig:runtimeapproxsqrt(deg)}, we consider different ranges of $\gamma$ values for each dataset, because of differences in which regimes are non-trivial. For each dataset and node-weighting, we first run \textsf{Nested} for a fine grid of $\gamma$ values to identify the smallest $\gamma$ at which at least one node is not deleted, and the largest value $\gamma_{U,\text{nest}}$ for which no nodes are deleted. Then we run a bisection approach using the LP method to find the largest parameter ($\gamma_{U,LP})$ for which the LP algorithm deletes no nodes. We set the lower limit of the grid to $\gamma_L/2$, and the upper limit to $\gamma_U = \max\{\gamma_{U,\text{Nest}}, \gamma_{U,LP} \}$ plus a small offset.
We use a log-spaced grid, where the number of grid points is $\lceil 20 \log_{10}(\gamma_U/\gamma_L) \rceil$. 


\FloatBarrier
\subsection{Effects of changing \texorpdfstring{$\gamma$}{gamma} and node weights}
We now turn to more qualitative case studies that explore how \SPECC{} identifies outliers in different datasets. This involves exploring the effect of changing $\gamma$ and node weights. In particular, we will consider differences between using unit weights for node deletion ($w(v) = 1$ for every $v$) and weights that are regularized in terms of node degree ($w(v) \in \{d_v, \sqrt{d_v} \}$).

\paragraph{Ranking nodes by varying $\gamma$ in \textsf{Nested} algorithm.}
Rather than choosing a single value of $\gamma$ and classifying nodes as either outliers or not (based on whether they are deleted), we can provide a more meaningful ranking of ``outlierness'' by varying $\gamma$. In more detail, we run our \textsf{Nested} algorithm for a range of $\gamma$ values, and rank nodes based on the largest value of $\gamma$ for which a node is deleted. 
If a node is deleted for very large values of $\gamma$, this provides a stronger indication that it should be viewed as an outlier, since it means the node is worth deleting even if the deletion cost is very high. Because we are using our \textsf{Nested} algorithm, we can be sure that if a node's rank is given by a certain large $\gamma$, then it will be deleted for all smaller (i.e., cheaper) node-deletion penalties as well. This nestedness property makes the ranking analysis more consistent and interpretable than running an algorithm that outputs solutions for different $\gamma$ values that are not directly related.

\begin{table}[t]
  \centering
  \caption{Top ten ranked outliers (ranked in terms of the largest $\gamma$ for which the node is still deleted) when using unit node weights and applying the \textsf{Nested} algorithm to our new Stackexchange Science dataset. The ranking is correlated with (but not perfectly monotonic in terms of) properties such that the degree (number of threads), color degree (number of sites user is active on), and number of conflicts. Identifying outliers for this dataset tends to highlight colors (i.e., sites) that are actually strongly related (e.g., philosophy/math/math history, CS/math/theoretical CS, etc.).}
  \label{tab:science}
  \small
  \setlength{\tabcolsep}{4pt}
  \begin{tabular}{lrrrrp{0.45\textwidth}r}
\toprule
Rank & Threads & Sites & Conflicts & Largest $\gamma$ & Sites posted in most (number of threads) & Other \\
\midrule
1 & 1,054 & 7 & 294,604 & 225 & Philosophy (600); Mathematics (399); History of Science and Mathematics (50) & 5 \\
2 & 877 & 10 & 195,225 & 201 & Computer Science (566); Mathematics (236); Theoretical Computer Science (50) & 25 \\
3 & 1,209 & 6 & 282,538 & 143 & Mathematics (911); Cross Validated (257); Economics (21) & 20 \\
4 & 481 & 6 & 67,199 & 127 & Chemistry (223); Mathematics (216); MathOverflow (17) & 25 \\
5 & 846 & 3 & 195,989 & 113 & Physics (495); Astronomy (268); Mathematics (83) & 0 \\
6 & 409 & 3 & 45,388 & 90.2 & Mathematics (233); Operations Research (146); MathOverflow (30) & 0 \\
7 & 220 & 3 & 14,967 & 80.5 & Cross Validated (103); Mathematics (81); Economics (36) & 0 \\
8 & 1,258 & 3 & 176,455 & 80.5 & Mathematics (1,098); MathOverflow (155); History of Science and Mathematics (5) & 0 \\
9 & 289 & 3 & 21,599 & 80.5 & Physics (175); Mathematics (97); Mathematics Educators (17) & 0 \\
10 & 368 & 3 & 29,347 & 80.5 & Physics (252); Mathematics (115); Computational Science (1) & 0 \\
\bottomrule
\end{tabular}
\end{table}

\paragraph{Case study on Science dataset.}
We begin with a case study on the Science dataset. Table~\ref{tab:science} displays the top 10 ranked nodes, based on the largest value of $\gamma$ for which the node is deleted. For this ranking, we use unit node weights ($w(v) = 1$). We observe that top-ranked users tend to be highly active across multiple sites covering closely related topics. High ranked nodes tend to be associated with higher values of degree, color degree, and the number of conflicts they occur in. At the same time, none of these characteristics by itself perfectly determines the ranking.
As another interesting takeaway, this list of top-ranked users highlights several sites (i.e., colors) which are very closely related to each other and should not be viewed as completely separate categories. For example, the highest-ranked user is most active on Philosophy, Mathematics, and History of Science and Mathematics, which clearly share substantial overlap. Similarly, other highly ranked users participate in related fields such as Computer Science, Mathematics, and Theoretical Computer Science. Thus, one potential use case of \SPECC{} is to help identify categorical features or labels in a dataset that are closely related or could in some cases be merged.

\paragraph{Case study on Cooking.}
We conduct a similar study on Cooking, where nodes are ingredients, hyperedges are recipes, and the color of a recipe indicates its cuisine (e.g., French, Mexican). Running \SPECC{} associates each ingredient with a specific cuisine, or marks it as an outlier that transcends cuisines. Table~\ref{tab:cooking-top10} shows that, under unit node-deletion weights, general-purpose ingredients such as salt and sugar, which clearly transcend cuisines, are deleted at larger values of $\gamma$. This captures one very intuitive notion of outliers. At the same time, giving all nodes the same deletion cost may bias the rankings in such a way that high-degree ingredients tend to get deleted more readily simply for being in a lot of recipes. This may be the case even if some ingredients are very popular but may still be naturally associated with one cuisine. Thus, we also compute the ranking using degree weights $w(v) = d_v$ (Table~\ref{tab:cooking-top10}, right). In this case, we begin to see slightly less generic (but still very general) ingredients more highly ranked.

\begin{table}[t]
  \caption{Cooking: the 10 ingredients the nested algorithm deletes longest under unit weights
    ($w(v)=1$) and degree weights ($w(v)=\deg(v)$). \textit{Recipes} is the number of recipes containing
    the ingredient and \textit{color-deg} is the number of cuisines it appears in. Several ingredients leave the
    deleted set at the same $\gamma$, and such ties are broken by node ID: under $w(v)=1$, ranks
    2--10 are tied with one further ingredient, and under $w(v)=\deg(v)$, ranks 7--10 are four
    of 63 tied ingredients.}
  \label{tab:cooking-top10}
  \setlength{\tabcolsep}{2pt}
  \begin{tabular}{r lrr | lrr}
    \toprule
    & \multicolumn{3}{c}{$w(v)=1$} & \multicolumn{3}{c}{$w(v)=\deg(v)$} \\
    \cmidrule(lr){2-4} \cmidrule(lr){5-7}
    \# & \textit{Ingredient} & \textit{Recipes} & \textit{color-deg} & \textit{Ingredient} & \textit{Recipes} & \textit{color-deg} \\
    \midrule
    1  & water             & 7,457  & 20 & ice cream                & 18    & 8  \\
    2  & sugar             & 6,434  & 20 & champagne                & 28    & 10 \\
    3  & garlic            & 7,380  & 20 & cold milk                & 22    & 11 \\
    4  & baking powder     & 1,738  & 20 & salted roast peanuts     & 33    & 5  \\
    5  & pepper            & 4,438  & 20 & chocolate bars           & 8     & 4  \\
    6  & milk              & 2,263  & 20 & stout                    & 26    & 7  \\
    7  & all-purpose flour & 4,632  & 20 & sugar                    & 6,434 & 20 \\
    8  & flour             & 1,348  & 20 & brussels sprouts         & 35    & 12 \\
    9  & butter            & 4,847  & 20 & Angostura bitters        & 13    & 8  \\
    10 & salt              & 18,048 & 20 & stir fry vegetable blend & 21    & 6  \\
    \bottomrule
  \end{tabular}
\end{table}
To refine this analysis further, Table~\ref{tab:cooking-rank-change}  reports ingredients whose ranking \emph{changes the most} between the unit-weight ranking and the degree-weight ranking. Interestingly, we find that many of these ingredients are high-degree nodes that have a comparatively high ranking under unit weights, even though they could arguably be said to belong primarily (even if not exclusively) to one specific cuisine. For example, salsa moves from rank 601 under unit weights to rank 3,208 under degree-based weights in our nested algorithm. Under unit weights, salsa is identified as an outlier relatively early because of its high degree. However, salsa is strongly associated with Mexican cuisine, making its retention within that category arguably more appropriate. Degree-based weighting better captures this distinction by reducing the tendency to classify high-degree, cuisine-relevant ingredients as outliers. Similarly, cajun seasoning and andouille sausage have relatively high ranks under unit-node weights, even though they have a very strong correlation with Cajun/Creole cuisine.
A similar observation holds for other ingredients in Table~\ref{tab:cooking-rank-change}.

\begin{table}[t]
  \caption{Cooking: the 10 ingredients whose rank under the nested algorithm changes most when
    the node weights change from $w(v)=1$ to $w(v)=\deg(v)$. Rank 1 is the ingredient deleted
    longest and 4,117 the one kept first; a positive change means the ingredient is kept earlier
    under degree weights. Tied ingredients share their average rank (48 tie groups under
    $w(v)=1$ and 47 under $w(v)=\deg(v)$). Rec.\ is the number of recipes; the main cuisine and
    its share are inferred from recipe counts and signature ingredients.}
  \label{tab:cooking-rank-change}
  \setlength{\tabcolsep}{2pt}
  \begin{tabular}{rlrlrr}
    \toprule
    \# & Ingredient & Rec. & Main cuisine & Rank $1 \to \deg$ & Change \\
    \midrule
    1  & salsa               & 963 & Mexican 96\%      & 601 $\to$ 3,208   & +2,607 \\
    2  & corn tortillas      & 965 & Mexican 98\%      & 1,185 $\to$ 3,708 & +2,523 \\
    3  & cajun seasoning     & 376 & Cajun/Creole 78\% & 601 $\to$ 2,842   & +2,242 \\
    4  & andouille sausage   & 267 & Cajun/Creole 84\% & 601 $\to$ 2,842   & +2,242 \\
    5  & shrimp paste        & 56  & Thai 52\%         & 1,544 $\to$ 3,744 & +2,200 \\
    6  & arborio rice        & 292 & Italian 89\%      & 862 $\to$ 3,045   & +2,182 \\
    7  & garam masala        & 925 & Indian 93\%       & 862 $\to$ 3,045   & +2,182 \\
    8  & tamarind pulp       & 22  & Indian 55\%       & 1,296 $\to$ 3,437 & +2,142 \\
    9  & fettucine           & 159 & Italian 86\%      & 732 $\to$ 2,842   & +2,110 \\
    10 & sherry wine vinegar & 41  & Spanish 39\%      & 732 $\to$ 2,842   & +2,110 \\
    \bottomrule
  \end{tabular}
\end{table}

\paragraph{Case study on Diseases.} We perform a similar ranking case study on Diseases. Outliers found by \SPECC{} tend to be generic symptoms that are not clearly indicative of any particular disease (e.g., nausea, high fever). Table~\ref{tab:diseases-top10} lists the ten symptoms that are deleted for the highest values of $\gamma$. Under unit weights, we observe that common symptoms, such as vomiting and skin rash, are identified as outliers relatively early, potentially due to their high degree across multiple diseases. When node weights are proportional to degree, some of these common symptoms, such as skin rash, are retained at larger values of $\gamma$, while somewhat less common symptoms emerge as outliers.

\begin{table}[t]
  \centering
  \caption{Diseases: the 10 symptoms the \textsf{Nested} algorithm deletes for the highest $\gamma$ values, under unit weights ($w(v)=1$) and degree
    weights ($w(v)=\deg(v)$). Rec.\ is the number of records containing the symptom and c-deg (color degree) is the number of
    diseases it appears in. Several symptoms leave the deleted set at the same $\gamma$, and such ties are broken by node ID: under $w(v)=1$, ranks 9--10 are tied with one further symptom, and under $w(v)=\deg(v)$, ranks 9--10 are tied with one further symptom.}
  \label{tab:diseases-top10}
  \setlength{\tabcolsep}{3pt}
  \begin{tabular}{rlrrlrr}
\toprule
 & \multicolumn{3}{c}{$w(v)=1$} & \multicolumn{3}{c}{$w(v)=\deg(v)$} \\
\cmidrule(lr){2-4}\cmidrule(lr){5-7}
\# & Symptom & Rec. & c-deg & Symptom & Rec. & c-deg \\
\midrule
1 & vomiting & 1,931 & 17 & skin rash & 793 & 7 \\
2 & skin rash & 793 & 7 & itching & 684 & 6 \\
3 & high fever & 1,374 & 12 & burning micturition & 218 & 2 \\
4 & fatigue & 1,949 & 17 & neck pain & 230 & 2 \\
5 & headache & 1,144 & 10 & swelling joints & 230 & 2 \\
6 & sweating & 684 & 6 & painful walking & 230 & 2 \\
7 & nausea & 1,156 & 10 & vomiting & 1,931 & 17 \\
8 & chest pain & 702 & 6 & joint pain & 690 & 6 \\
9 & yellowish skin & 920 & 8 & chest pain & 702 & 6 \\
10 & loss of appetite & 1,162 & 10 & dizziness & 339 & 3 \\
\bottomrule
\end{tabular}
\end{table}

\FloatBarrier
\section{Conclusions}
This paper introduces a generalization of edge-colored clustering called \SPECC{} that handles node outliers by including a soft penalty for deleting nodes. This formulation is amenable to constant-factor approximation algorithms. In particular, we develop an LP-based $(2-2/k)$-approximation based on a reduction to node-weighted multiway cut. We also introduce the notion of a \emph{nested} approximation algorithm, which solves a sequence of \SPECC{} problems with increasing node-deletion penalties while ensuring that the node (and edge) deletion sets are monotonically shrinking (or growing) in terms of the parameter $\gamma$. In practice, our methods allow us to identify meaningful outliers in various types of categorical clustering applications. Our experiments on the new Stackexchange Science dataset highlight that categories can be closely related or overlapping in some applications. During the course of our research we also identified many datasets that are made up of multiway interactions with more than one category. One possibly interesting direction for future research is to develop alternate notions of ECC that allow hyperedges to have multiple colors. 

\section*{Acknowledgments}
This material is based upon work supported by the U.S. Department of Energy, Office of Science, Office of Advanced Scientific Computing Research under the project RAMSHORN: Randomized Algorithms for Massive, Scientific, Higher-Order Relational Networks, Award Number DE-SCL0000093. The research is also funded in part by the Army Research Office under ARO award \#W911NF‐24-1-0156. The views and conclusions contained in this document are those of the authors and should not
be interpreted as representing the official policies, either expressed or implied, of the Army Research Office
or the U.S. Government.

\section*{AI disclosure statement}
AI tools were used at different points to help generate TikZ code for illustrative figures, help implement algorithms in Julia, plots results, format tables, and build the new Stackexchange hypergraph. The authors also used AI tools to help design and prove approximation guarantees for Algorithms~\ref{alg:direct-lp-rounding} and~\ref{alg:better-direct-lp-rounding}. This involved prompting an LLM to generate ideas, which were then iteratively checked and refined by the authors. For the manuscript write-up, LLMs were used sparingly. Most sections were drafted and edited entirely by the authors. LLMs assisted in initially drafting some technical pieces (e.g., the approximation proof for Algorithm~\ref{alg:better-direct-lp-rounding}) which were then heavily checked and edited by the authors. Authors take responsibility for all aspects of the work.

\bibliographystyle{plain}
\bibliography{refs}
\appendix

\end{document}